\documentclass[useAMS,usenatbib]{mn2e}

\usepackage{acronym}
\usepackage{amsmath}
\usepackage{gensymb}
\usepackage{graphicx}
\usepackage{hyperref}
\usepackage{textcomp}
\usepackage{times}
\usepackage{cleveref}

\newcommand{\cbc}{\mathrm{BNS}}
\newcommand{\dur}{\mathrm{dur}}
\newcommand{\fov}{\mathrm{FOV}}
\newcommand{\sint}{\mathrm{int}}
\newcommand{\iso}{\mathrm{iso}}
\newcommand{\jet}{\mathrm{jet}}
\newcommand{\obs}{\mathrm{obs}}
\newcommand{\peak}{\mathrm{peak}}
\newcommand{\sref}{\mathrm{ref}}
\newcommand{\thermal}{\mathrm{th}}
\newcommand{\thres}{{\ast}}

\newcommand{\given}{\,\,|\,}
\newcommand{\diff}{\, \mathrm{d}}
\newcommand{\planck}{\textit{Planck}}

\title[Radio Afterglows from Compact Binary Coalescence]
      {Detectability of Late-Time Radio Afterglows from Compact Binary Coalescence}
\author[L. Feng et al.]
       {L.~Feng,$^{1}$\thanks{E-mail: lufeng@mit.edu} R.~Vaulin,$^{1}$ and J.~N.~Hewitt$^{1}$ \\
         $^{1}$Kavli Institute for Astrophysics and Space Research, 
         Massachusetts Institute of Technology, Cambridge, MA 02139, USA}

\begin{document}
\maketitle
\label{firstpage}

\begin{abstract}
Electromagnetic (EM) follow-up of gravitational wave (GW) candidates is important for verifying their astrophysical nature and studying their physical properties. While the next generation of GW detectors will have improved sensitivities to make the first detection of GW events, their ability to localize these events will remain poor during the early days of their operation. This makes EM follow-up challenging for most telescopes. Many new low frequency radio instruments have come online recently or will come online over the next few years, and their wide fields of view allow them to cover large areas of the sky in a short amount of time. This paper studies comprehensively the detectability of radio afterglows from compact binary coalescence (CBC), a predicted GW source and the most promising progenitor of short gamma-ray bursts. We explore the properties of simulated afterglow lightcurves from the forward shock for a range of source and observer parameters, then we use these lightcurves to estimate the expected rates of detection for different radio instruments and survey methods. Detecting radio afterglows and constraining their properties and rates are feasible with the current and upcoming widefield radio instruments. As a result, widefield radio instruments will play an important role in the EM follow-up of GW events. 
\end{abstract}


\section{Introduction}\label{introduction}

The coalescence of two compact objects, e.g. \ac{BNS} or a neutron star and a black hole, are predicted sources of \ac{EM} and \ac{GW} emission. Joint \ac{EM} and \ac{GW} observations of these systems are complementary as they probe different physical processes and are necessary for certain science objectives (e.g. \citealt{bloom2009}, \citealt{phinney2009}, \citealt{ligo2012}). For example, \ac{EM} detections of \ac{GW} events will measure the source redshift and break the degeneracy between the source distance and its inclination angle. This will improve estimations of astrophysical parameters such as the Hubble parameter since \ac{GW} detectors have different systematic uncertainties (e.g. \citealt{schutz1986}, \citealt{dalal2006}, \citealt{nissanke2010}). Furthermore, \ac{GW} measurements of the inclination angles will improve our understanding of the dynamics and energetics of the \ac{EM} counterparts \citep{arun2014}. 

The next generation of \ac{GW} detectors, such as Advanced LIGO (aLIGO, \citealt{harry2010}) and Advanced Virgo \citep{virgo2009}, will be coming online in 2015 with improved sensitivity to detect \ac{GW} events directly for the first time. \ac{EM} follow-up of \ac{GW} candidates will be important for confirming the astrophysical nature of these events. Many \ac{EM} counterparts have been proposed, including kilonovae, \ac{SGRB}, and afterglows (e.g. \citealt{eichler1989}, \citealt{narayan1992}, \citealt{li1998}, \citealt{metzger2010}, \citealt{metzger2012}, \citealt{piran2013}). To date, observational evidence supporting the connection between these \ac{EM} counterparts and \ac{CBC} remains indirect or uncertain, for instance the diverse properties of \ac{SGRB} host galaxies (\citealt{berger2009}, \citealt{fong2010}) and one possible kilonova association with a \ac{SGRB} (\citealt{tanvir2013}, \citealt{berger2013}). Coincident detections of these \ac{EM} counterparts and \ac{GW} emission would firmly establish the origin of these events. However, the sky localization of \ac{GW} candidates will be poor during the early days of \ac{GW} detector operation, ranging from 100--1000\,deg$^{2}$ (\citealt{ligo2013}, \citealt{singer2014}). This presents a challenge for \ac{EM} follow-up with most telescopes as they have much smaller fields of view in comparison. 

Many new widefield radio instruments have recently begun operation or will soon begin operation: the Long Wavelength Array\footnote{http://lwa.phys.unm.edu/} Station 1 (LWA1, \citealt{ellingson2013}), the Low-Frequency Array\footnote{http://www.lofar.org/} (LOFAR, \citealt{vanhaarlem2013}), the Murchison Widefield Array\footnote{http://mwatelescope.org/} (MWA, \citealt{tingay2013}), the Canadian Hydrogen Intensity Mapping Experiment\footnote{http://chime.phas.ubc.ca/} (CHIME), and the Australian Square Kilometer Array Pathfinder\footnote{http://www.atnf.csiro.au/projects/askap/} (ASKAP). More instruments are planned for the future, such as the Hydrogen Epoch of Reionization Array\footnote{http://reionization.org/} (HERA) and the Square Kilometer Array\footnote{http://www.skatelescope.org/} (SKA). The large fields of view of these instruments (30--600\,deg$^{2}$) make \ac{EM} follow-up of \ac{GW} candidates feasible. 

Most of these instruments operate at low frequencies ($<500$\,MHz), where the expected \ac{EM} counterpart of a \ac{GW} event is a \ac{SGRB} afterglow. So far, there have been no detections of \ac{SGRB} afterglows at low frequencies and only 3 detections at high frequencies ($>5$\,GHz): GRB~050724, GRB~051221A, and GRB~130603B (\citealt{berger2005}, \citealt{soderberg2006}, \citealt{fong2014}). This is not surprising given the sample of radio afterglows discussed in \citet{chandra2012}. Few \ac{SGRB}s, if any, have been observed at low frequencies. \ac{SGRB}s are also intrinsically fainter than long GRBs, releasing less energy in total and occurring in a less dense medium. Radio emission from afterglows usually peak on time-scales of weeks to months, if not longer, and few \ac{SGRB}s have been observed on this time-scale. Furthermore, \ac{SGRB}s triggered by $\gamma$-rays have been cosmological ($z > 0.1$) if they have measured redshifts at all. In contrast, detectable \ac{GW} events will be nearby ($z < 0.1$). For these events, radio afterglows are still expected to be faint and long-lasting (\citealt{nakar2011}, \citealt{metzger2012}, \citealt{kelley2013}), but this paper shows that, within a plausible range of afterglow model parameters, there is a spread in the distributions of peak fluxes and durations of these afterglows, suggesting that a subset of afterglows could be detectable by the new widefield radio instruments. However, the results are sensitive to many model parameters that are still currently uncertain. 

Distinguishing faint \ac{SGRB} afterglows from other slow transients such as radio supernovae could be an additional challenge. However, the radio transient sky at low frequencies is poorly understood. Many radio transients are expected to exist \citep{cordes2004}, but few have been detected so far \citep{frail2012}. While previous transient surveys at low frequencies were limited by sensitivity (\citealt{lazio2010}, \citealt{bell2014}, Kudryavtseva et al. in prep) or field of view \citep{jaeger2012}, surveys with these new instruments will be able to characterize the expected rate of background radio transients for \ac{EM} follow-up before the advanced \ac{GW} detectors turn on. If \ac{SGRB}s are indeed associated with \ac{CBC}, these instruments can also search for both on-axis and off-axis afterglows to constrain the \ac{CBC} rate, which is uncertain by three orders of magnitude \citep{abadie2010}. 

Previous radio searches for orphan afterglows have yielded null results (\citealt{levinson2002}, \citealt{galyam2006}) but at relatively limited sensitivity (6\,mJy). As this paper shows, the new widefield radio instruments have good thermal sensitivities in addition to large fields of view, making them suitable for transient surveys and follow-up observations. This work is complementary to recent studies of radio emission from subrelativistic outflows of \ac{CBC} \citep{piran2013} in addition to the detectability of radio afterglows of long GRBs at high frequencies (\citealt{ghirlanda2013}, \citealt{ghirlanda2014}) and high redshifts \citep{zhang2014}. We explore the properties of simulated \ac{SGRB} afterglow lightcurves at radio frequencies for a range of source and observer parameters (Section~\ref{lightcurves}) and characterize the detectability of these events for radio instruments (Section~\ref{detectability}). Then we estimate the expected rates of detection for different instruments and survey methods (Section~\ref{rates}). We expect most of these instruments will be able to detect \ac{SGRB} afterglows or constrain optimistic \ac{CBC} models and compare our results to recent work by others in Section~\ref{discussion}. While recent radio observations of GRB~130427A show that there is bright ($\sim$\,mJy) radio emission due to the reverse shock at early times \citep{anderson2014}, this emission component is not included in this paper, which only considers late-time ($>1$\,d) afterglow emission from the forward shock, but it will be subject to future studies.

\section{Lightcurve Properties}\label{lightcurves}

The afterglow emission of a \ac{SGRB} is synchrotron radiation produced when the relativistic ejecta creates a shock in the surrounding medium (see \citealt{nakar2007} and \citealt{berger2013_review} for recent reviews). The shape of the lightcurve depends on the properties of the burst, the microphysics of synchrotron radiation, and the parameters specifying an observer (\citealt{sari1998}, \citealt{granot2002}). Observationally, \ac{SGRB}s have isotropic energies $10^{49} \la E_{\iso} \la 10^{51}$\,ergs (\citealt{nakar2007}, \citealt{berger2013}, and references therein). Their jet opening angles are difficult to measure and thus have large uncertainties, but a few jet break measurements suggest $\theta_{\jet} \sim 10\degr$ (\citealt{burrows2006}, \citealt{soderberg2006}, \citealt{fong2012}, \citealt{fong2014}). Their circumburst environments generally have low inferred densities $10^{-5} \la n \la 1$\,cm$^{-3}$ (\citealt{soderberg2006}, \citealt{panaitescu2006}, \citealt{fong2011}, \citealt{fong2012}, \citealt{fong2014}), consistent with the expectations for \ac{BNS} mergers \citep{perna2002}. The results from these observations motivate the parameter space that we explore in this paper. 

Recently, \citet{vaneerten2012_boxfit} developed a numerical tool \textsc{boxfit} that generates afterglow lightcurves quickly for arbitrary burst and observer parameters. The availability of this tool has allowed us to improve on the previous estimates of \ac{SGRB} afterglow properties derived from analytical approximations (\citealt{nakar2011}, \citealt{metzger2012}). \textsc{boxfit} calculates the fluid state of the shock by interpolating the results of two-dimensional relativistic hydrodynamics jet simulations after applying the analytical solutions of \citet{blandford1976} to the ultra-relativistic phase of the shock expansion. Then it calculates the lightcurve by solving the linear radiative transfer equations for synchrotron radiation. 

Using this tool, we generate lightcurves of \ac{SGRB} afterglows to study their properties and detectability at radio frequencies. We specify \textsc{boxfit} to use the Blandford--McKee solutions for $200 > \gamma > 25$ where $\gamma$ is the Lorentz factor of the fluid directly behind the shock front. We also fix the parameters describing the microphysics of synchrotron radiation to their characteristic values (e.g. \citealt{vaneerten2011}, \citealt{piran2013}), listed in \Cref{tb:simulation_parameters} along with the other relevant parameters, which are consistent with observations. We explore a range of energies and densities corresponding to known constraints and expectations. For each combination of $E_{\iso}$ and $n$, we generate lightcurves at 4 observer frequencies sampling the range covered by widefield radio instruments and at 11 observer angles spaced linearly between $0\degr$ (on-axis) and $90\degr$ (off-axis). Each lightcurve consists of 350 time samples spaced logarithmically between $0.1$ and $10^{6}$\,d, capturing the evolution of the afterglow from early to late times. The bursts are located at $10^{27}$\,cm ($324$\,Mpc), a distance comparable to the average aLIGO \ac{BNS} range at design sensitivity \citep{ligo2013} but is otherwise an arbitrary choice, and the lightcurves are generated in the source frame. 

\begin{table*}
  \centering
  \begin{minipage}{105mm}
  \caption{Parameters used to generate afterglow lightcurves.}
  \begin{tabular}{lll} 
    \hline
    Observer Parameters & Burst Parameters\footnote{$E_{\jet}$ is related to $E_{\iso}$ through $E_{\iso} = 2E_{\jet}/\theta_{\jet}^{2}$ for $\theta_{\jet} \ll 1$.} & Microphysics\footnote{$\xi_{N}$ is the fraction of accelerated electrons, $p$ is the power-law slope of the electron energy distribution, $\epsilon_{e}$ is the fraction of internal energy in electrons, and $\epsilon_{B}$ is the fraction of internal energy in the magnetic field.} \\
    \hline
    $\nu_{\obs} = 60, 150, 600, 1430$\,MHz & $\theta_{\jet} = 11.5\degr$ (0.2\,rad) & $\xi_{N} = 1.0$ \\
    $\theta_{\obs} = 0\degr$--$90\degr$ & $E_{\jet} = 10^{48}, 10^{50}$\,ergs & $p = 2.5$ \\
    $t_{\obs} = 0.1$--$10^{6}$\,d & $E_{\iso} = 5 \times 10^{49}, 5 \times 10^{51}$\,ergs & $\epsilon_{e} = 0.1$ \\
    $d_{\sref} = 10^{27}$\,cm $(z = 0)$ & $n = 10^{-5}, 10^{-3}, 1.0$\,cm$^{-3}$ & $\epsilon_{B} = 0.1$ \\
    \hline
  \end{tabular} 
  \label{tb:simulation_parameters}
  \end{minipage}
\end{table*}

\begin{figure}
  \centering
  \includegraphics[width=0.5\textwidth]{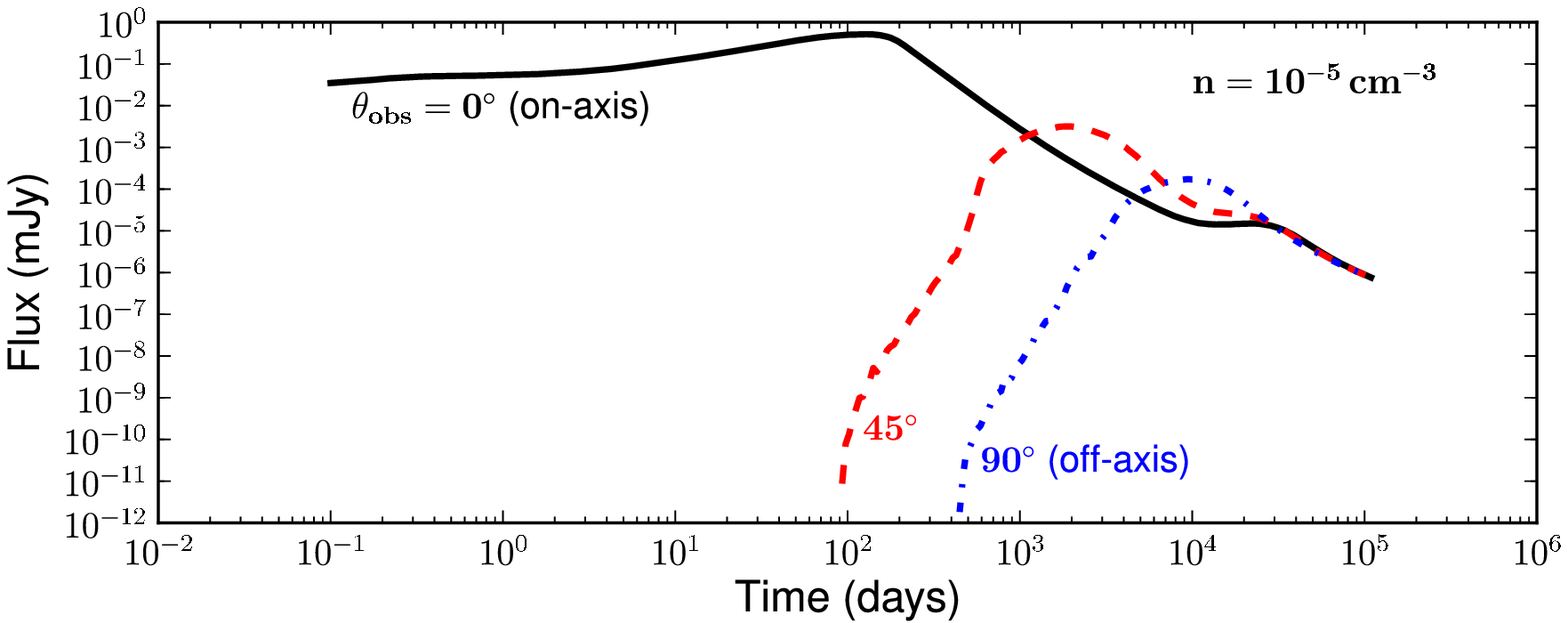}
  \includegraphics[width=0.5\textwidth]{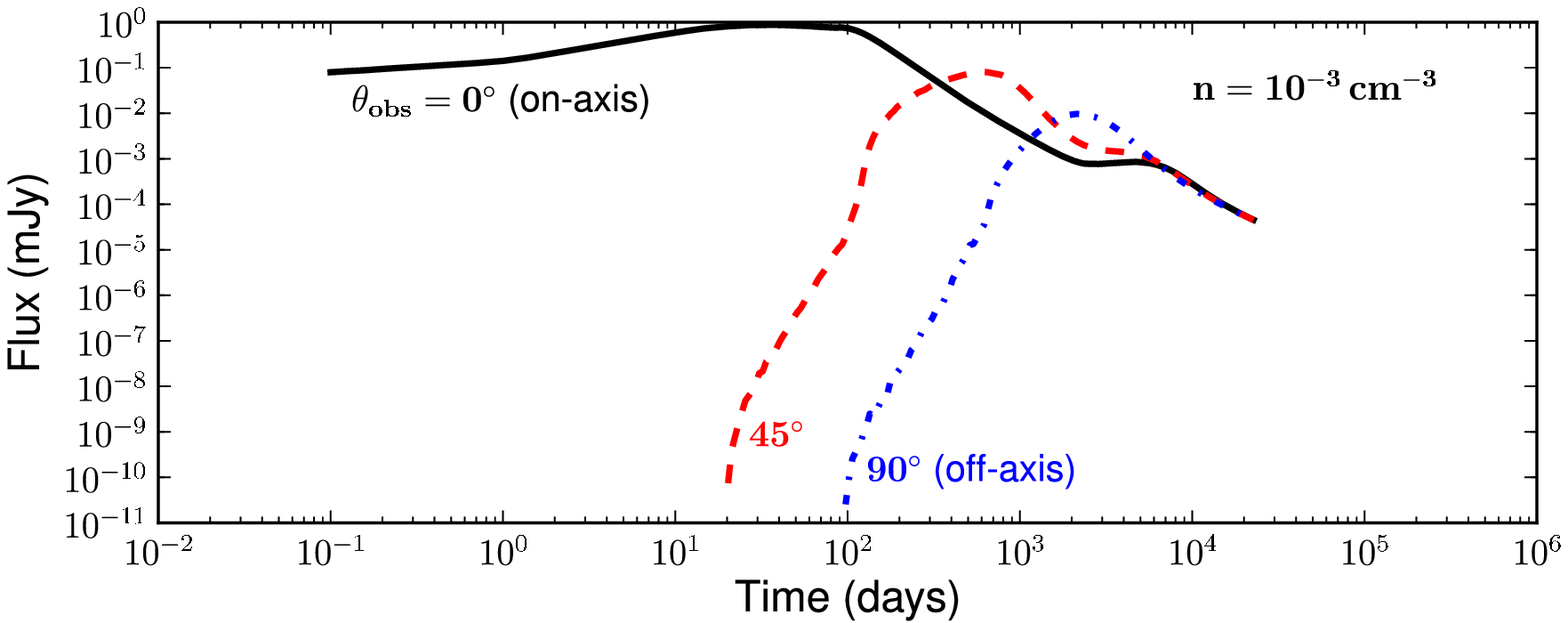}
  \includegraphics[width=0.5\textwidth]{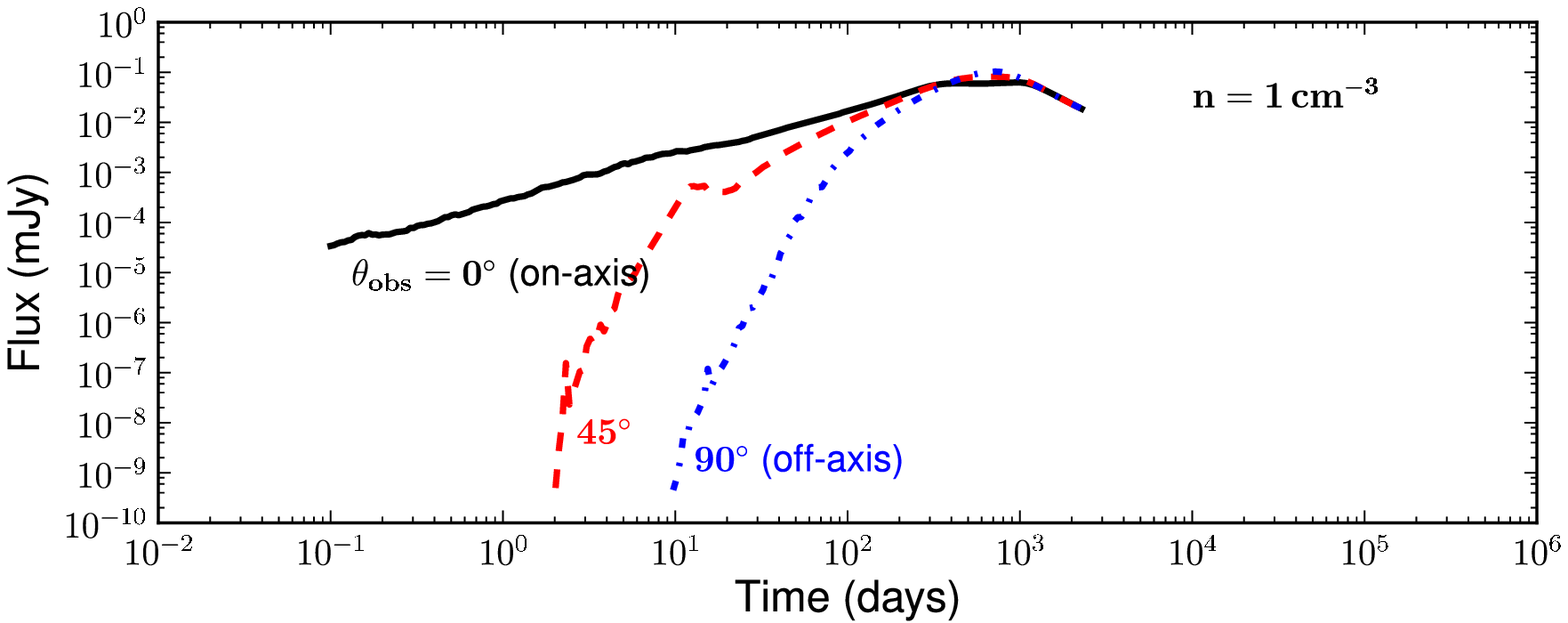}
  \caption{Examples of simulated afterglow lightcurves at 150\,MHz for different values of $n$ and $\theta_{\obs}$. These bursts are located at 324\,Mpc with $E_{\iso} = 5 \times 10^{51}$\,ergs and $\theta_{\jet} = 11.5\degr$. Early-time emission is brighter for on-axis observers because of collimated outflows and relativistic beaming whereas late-time emission is isotropic. The late-time bump is caused by the counter-jet. The lightcurves in the bottom panel are qualitatively different from the ones in the top two panels because synchrotron self-absorption is stronger at higher densities.}
  \label{fig:lightcurve_examples}
\end{figure}

Although the detailed shape of an afterglow lightcurve depends on many parameters, the general shape rises and falls on time-scales of months to years (\Cref{fig:lightcurve_examples}). At early times, an on-axis observer sees more emission than an off-axis observer because of collimated outflows and relativistic beaming. At late times, the on-axis and off-axis lightcurves become indistinguishable as the emission becomes isotropic. The counter-jet contributes to a late-time brightening of the lightcurve, an effect that is most prominant for an on-axis observer. While synchrotron emission becomes stronger as $n$ increases, synchrotron self-absorption becomes even stronger at low radio frequencies. In our lightcurves, synchrotron self-absorption is most prominent when $n = 1$\,cm$^{-3}$, causing a much slower rise in flux at early times. 

To capture the properties of an ensemble of afterglow lightcurves, we generate a sample of bursts that is uniformly distributed in energy ($5 \times 10^{49} \le E_{\iso} \le 5 \times 10^{51}$\,ergs), jet orientation ($-1 \le \cos \theta_{\obs} \le 1$), and comoving volume ($z \le 1$, chosen because bursts with larger distances are unlikely to be detectable by instruments operating in the next decade). Instead of rerunning \textsc{boxfit} with new parameters, we use the analytical energy-flux scaling relation derived by \citet{vaneerten2012} to determine the peak fluxes of the lightcurves over a continuous range of energies: $E_{\iso}' = \kappa E_{\iso}$ and $f_{\peak}' = \kappa f_{\peak}$ where $\kappa$ is a scaling parameter for a fixed density, distance, and observer angle. Then we scale these fluxes according to their luminosity distances \citep{hogg1999}: $f_{\peak}(d_{\mathrm{L}}) = (1+z)f_{\peak}'(d_{\sref} / d_{\mathrm{L}})^{2}$. We also scale the durations of these lightcurves according to $t_{\dur}' = \kappa^{1/3} t_{\dur}$ \citep{vaneerten2012}, where we define $t_{\dur}$ to be the time during which flux $> 0.5f_{\peak}$. 

\begin{figure}
  \centering
  \includegraphics[width=0.5\textwidth]{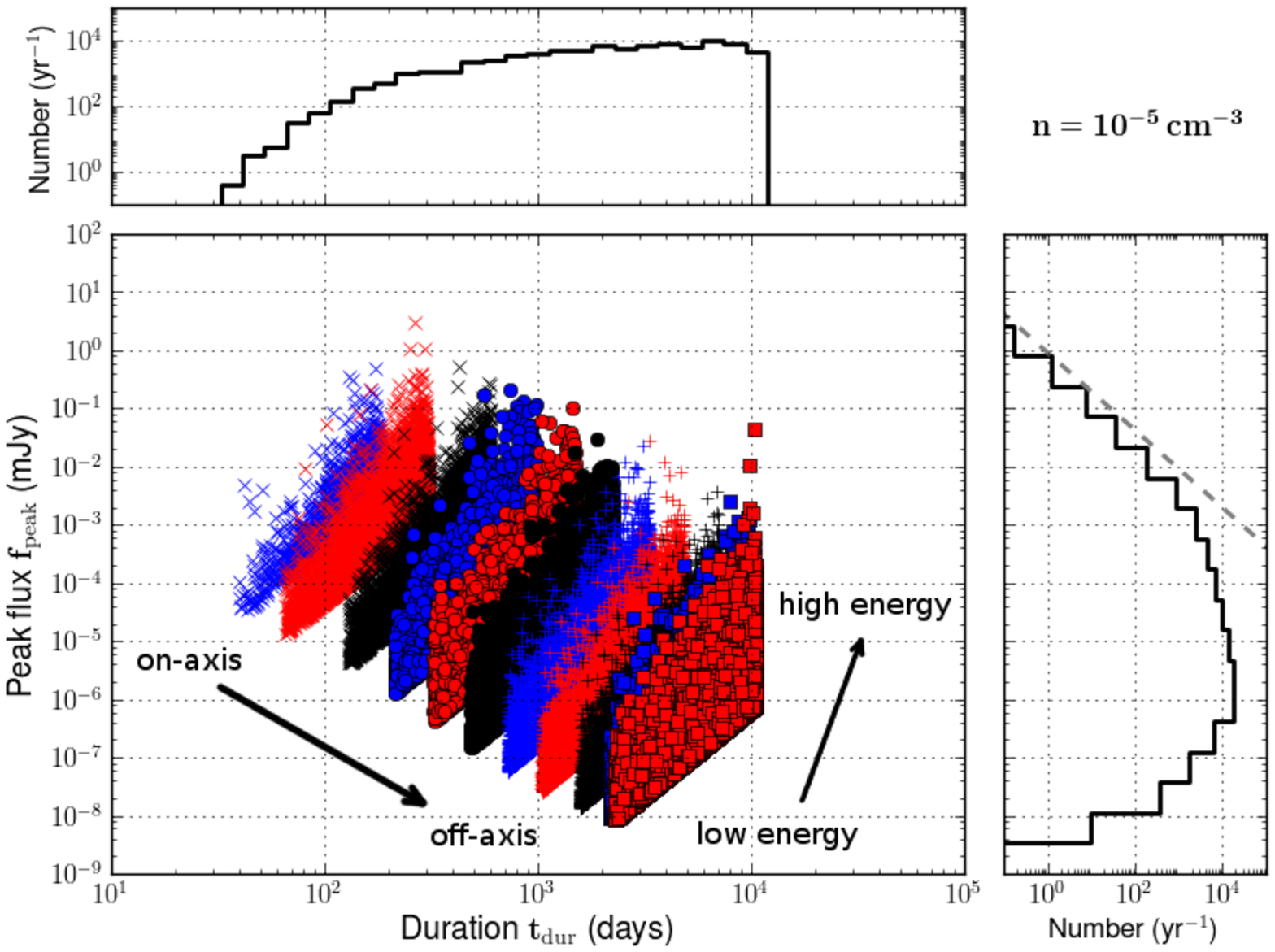}
  \includegraphics[width=0.5\textwidth]{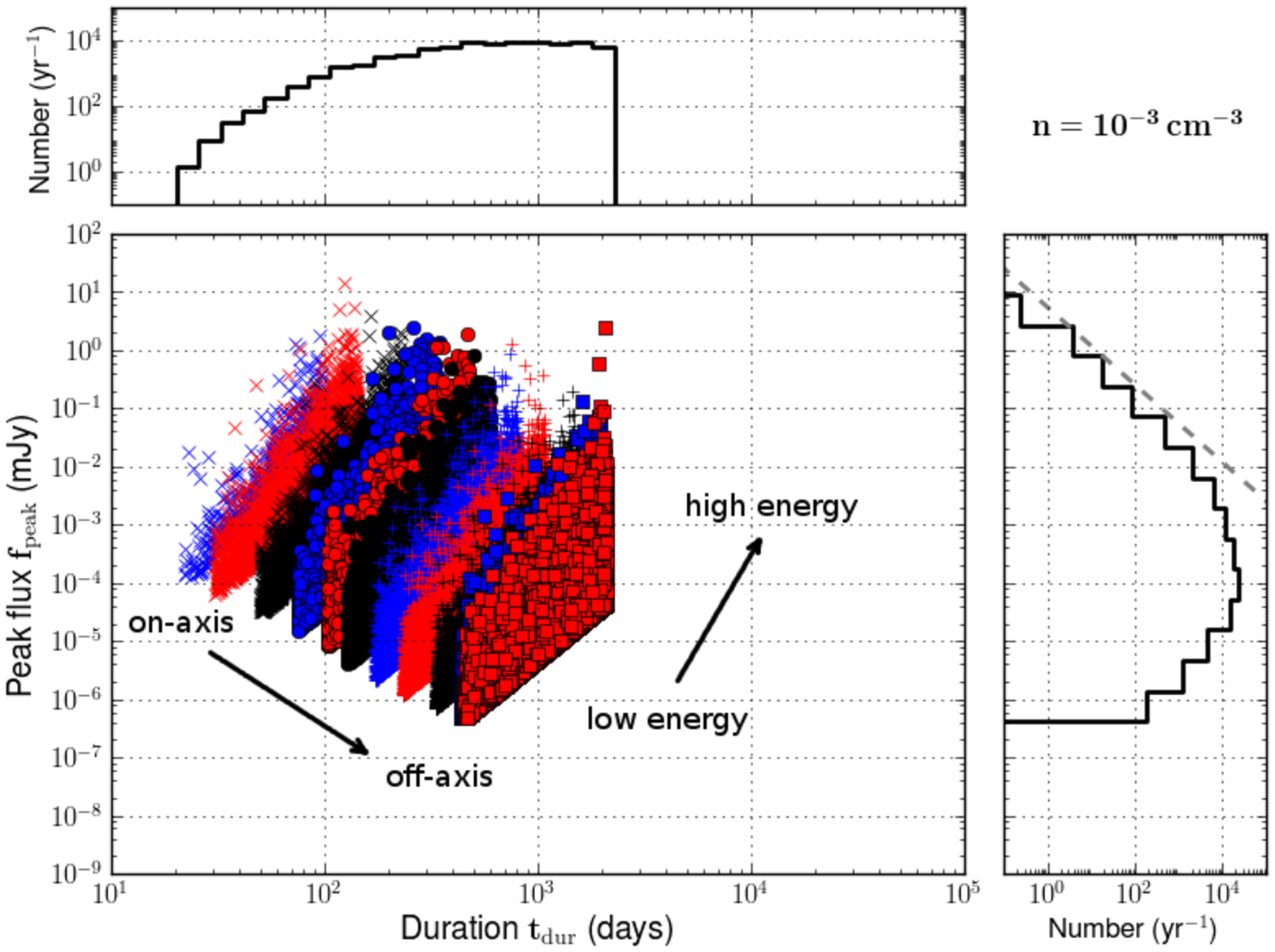}
  \includegraphics[width=0.5\textwidth]{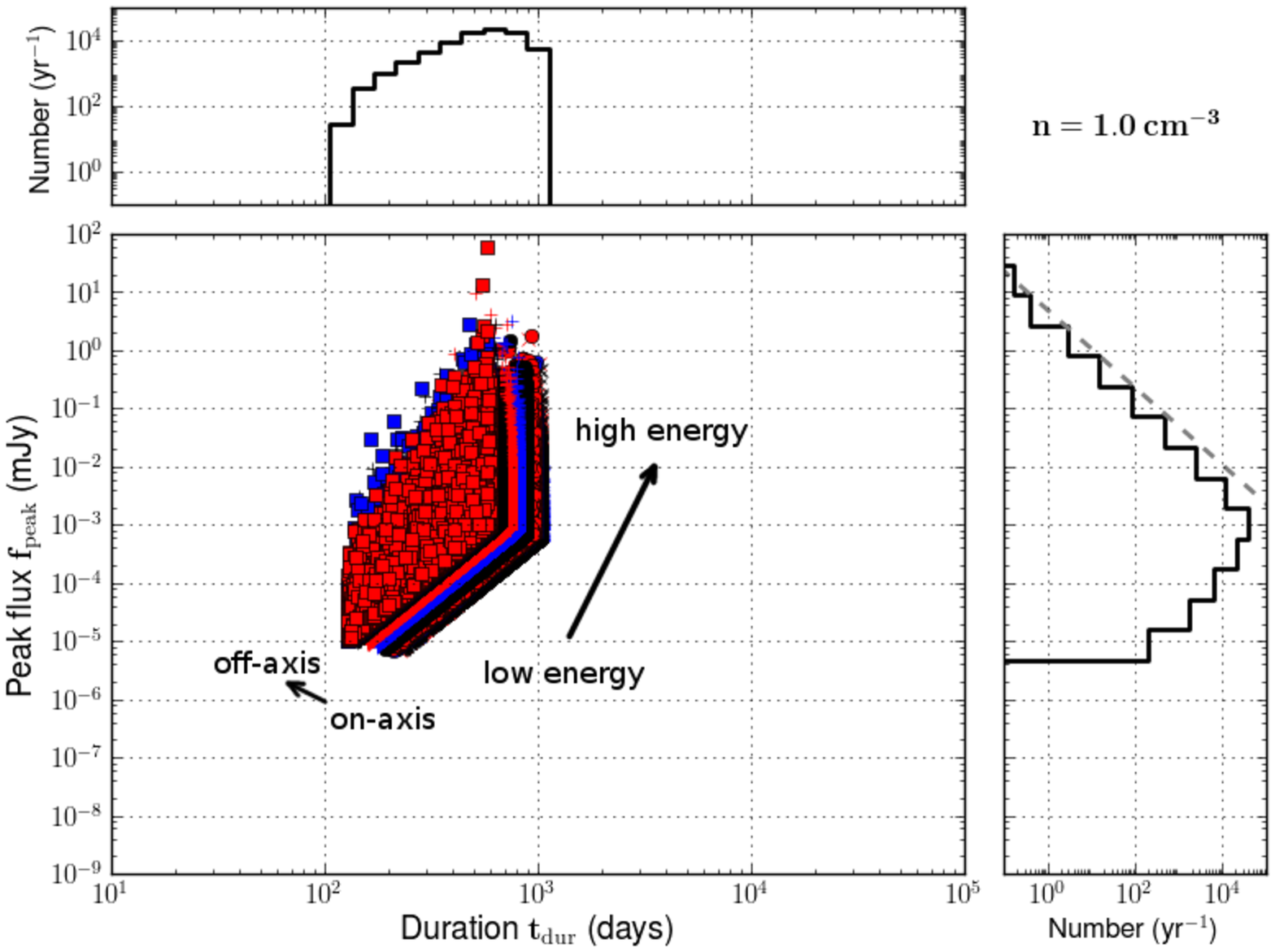}
  \caption{Distributions of $f_{\peak}$ and $t_{\dur}$ for the simulated afterglow lightcurves in a volume-limited sample at 150\,MHz and different values of $n$. All distributions in this paper are normalized to the realistic \ac{CBC} rate of 1\,Mpc$^{-3}$\,Myr$^{-1}$ for $2\pi$ sky area and will not be mentioned separately. The sample is uniformly distributed in $E_{\iso}$, jet orientation, and volume. The distinct clusters in the scatter plot are caused by binning in $\theta_{\obs}$. The vertical edges are caused by the energy cutoffs at $5 \times 10^{49}$ and $5 \times 10^{51}$\,ergs. The diagonal edges are caused by the distance cutoff at $z = 1$. The dashed line in the histogram panels on the right is a reference line with slope $-3/2$ for $N \propto f^{-3/2}$. See text for more details.}
  \label{fig:flux_duration_distribution}
\end{figure}

\begin{figure}
  \centering
  \includegraphics[width=0.5\textwidth]{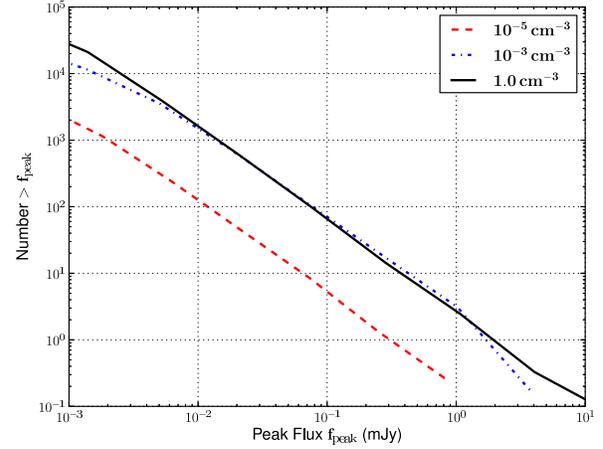}
  \caption{Cumulative distributions of $f_{\peak}$ for the simulated afterglow lightcurves at 150\,MHz. The sample is the same as that of \Cref{fig:flux_duration_distribution} but plotted for $f_{\peak} > 1$\,$\umu$Jy. This shows the population of afterglows that could be detected as sources given a particular instrument flux sensitivity at 150\,MHz.}
  \label{fig:cumulative_flux_intrinsic}
\end{figure}

As evident from the distributions of $f_{\peak}$ and $t_{\dur}$ shown in \Cref{fig:flux_duration_distribution}, most bursts are faint ($\la \umu$Jy) and long-lasting ($\ga$\,yr), confirming the results of previous studies. However, there is a spread, implying that there may be bursts detectable with current widefield radio instruments. The spread is dependent on the model parameters, which are fairly uncertain. This is also evident in the cumulative distributions of $f_{\peak}$ for the same sample (\Cref{fig:cumulative_flux_intrinsic}).

\Cref{fig:flux_duration_distribution} also shows that bursts become brighter and longer-lasting as their energies increase, but they become fainter and longer-lasting when they are more off-axis. This is consistent with the results from \citet{kelley2013}, who considered only the lowest energy bursts. The trend along observer angle is absent for $n = 1$\,cm$^{-3}$ because emission is isotropic by the time the system becomes optically thin. The two outlier points (red squares) just happen to be nearby samples. These properties suggest that detectable bursts will be more on-axis, have higher energies, and occur in higher density environments. Observing these bursts will take $\sim 1$\,yr, which can be undertaken by a realistic survey. 

A recent theoretical study of late-time afterglow lightcurves by \citet{sironi2013} argues that these afterglows could be a factor of a few brighter than previously expected if the bulk of the shock-accelerated electrons are non-relativistic, which could improve the detectability of these events, but this effect is not included in our study.

\section{Detection Metric}\label{detectability}

The detectability of radio afterglows depends not only on the intrinsic properties of \ac{SGRB}s, as shown in Section~\ref{lightcurves}, but also on the sensitivity of the radio instrument. The simplest characterization of the sensitivity of a radio interferometer is the thermal noise $\sigma_{\thermal}$, or point source sensitivity: 
\begin{equation}\label{eq:thermal_noise}
  \sigma_{\thermal} = \left( \frac{2k_{\mathrm{B}} T_{\mathrm{sys}}}{A_{\mathrm{eff}} N_{\mathrm{ant}} \epsilon_{c}} \right) \frac{1}{\sqrt{N_{\mathrm{pol}}B\,t_{\sint}}}
\end{equation}
where $k_{\mathrm{B}}$ is the Boltzmann constant, $T_{\mathrm{sys}}$ is the system temperature, $A_{\mathrm{eff}}$ is the effective area of each antenna, $N_{\mathrm{ant}}$ is the number of antennas, $\epsilon_{c}$ is the correlator efficiency, $N_{\mathrm{pol}}$ is the number of polarizations, $B$ is the instantaneous bandwidth, and $t_{\sint}$ is the snapshot integration time. \Cref{tb:telescope_parameters} lists the sensitivities and other relevant system parameters for the radio instruments selected for our analysis. 

\begin{table*}
  \centering
  \begin{minipage}{120mm}
  \caption{System parameters for present and future widefield radio instruments. $\nu_{\obs}$ is the observer frequency at which we generate the afterglow lightcurves. $\sigma_{\mathrm{1h}}$ is the thermal sensitivity of a 1-hour snapshot. $\Omega_{\fov}$ is the effective field of view for this snapshot. Except for CHIME and HERA, we scale $\sigma_{\mathrm{1h}}$ from the published thermal sensitivity for each instrument according to \Cref{eq:thermal_noise}. $^{\ast}$For CHIME and HERA, we calculate an average $\sigma_{\thermal}$ of a 1-day drift-scan snapshot (see Appendix~\ref{appendix_chime_sens}). $^{\dagger}$Both LWA1 and LOFAR can form multiple beams simultaneously, increasing their sky coverage by decreasing their bandwidth (see \citealt{ellingson2013} and \citealt{vanhaarlem2013}), and we choose a large $\Omega_{\fov}$ for them.}
  \begin{tabular}{lccccc} 
    \hline
    Instrument & Frequency Range & Bandwidth & $\nu_{\obs}$ & $\sigma_{\mathrm{1h}}$ & $\Omega_{\fov}$ \\ 
    & (MHz) & (MHz) & (MHz) & (mJy) & (deg$^{2}$) \\ 
    \hline
    LWA1$^{\dagger}$ & 10--88 & 16 & 60 & 16.8 & $4 \times 61$ \\
    LOFAR Low$^{\dagger}$ & 10--80 & 3.66 & 60 & 17.5 & $48 \times 74.99$ \\
    LOFAR High$^{\dagger}$ & 110--240 & 3.66 & 150 & 0.877 & $48 \times 11.35$ \\
    MWA & 80--300 & 30.72 & 150 & 0.913 & 610 \\
    CHIME Pathfinder$^{\ast}$ & 400--800 & 400 & 600 & 0.240 & 20626 \\
    CHIME$^{\ast}$ & 400--800 & 400 & 600 & 0.036 & 20626 \\
    ASKAP & 700--1800 & 300 & 1430 & 0.029 & 30 \\ 
    HERA$^{\ast}$ & 50--225 & 100 & 150 & 0.017 & 2712 \\ 
    SKA1 Low & 50--350 & 250 & 150 & 0.002 & 27 \\ 
    \hline
  \end{tabular}
  \label{tb:telescope_parameters}
  \end{minipage}
\end{table*}

As afterglows are faint and slow transients, their robust detection requires an application of the time-series data analysis techniques adopted and optimized for dealing with radio data. This section discusses the primary sources of noise in such measurements and outlines a possible search algorithm. Based on this algorithm, we derive the criteria for the detectability of afterglows used in the rest of the paper. The full treatment of the detection problem requires the incorporation of the specific characteristics of the instrument (e.g. calibration errors, incomplete $uv$-coverage, ionospheric fluctuations, observing mode, etc.) and is beyond the scope of this paper. Here we consider a simplified version of the analysis, which none the less allows us to establish a realistic criteria for afterglow detection. 

The fundamental sources of noise in the radio searches for transient signals are the thermal noise, from the antennas and the radio sky, and the classical and sidelobe confusion noise, from the unresolved constant point sources \citep{condon1974}. Depending on the angular resolution of the telescope, the confusion noise could be above or below the thermal noise. Whatever might be the case, the resolved constant sources, especially if they are dim, could be confused with or cover up a transient source, and thus should be considered as a component of the confusion noise. Uncertainty in calibration and incomplete $uv$-coverage may contribute as additional sources of non-stationary noise, and, in some cases, may dominate the noise budget. In the real search, the noise of the telescope should be carefully characterized to account for all of these factors. In what follows though, we neglect all other contributions to the noise except for the thermal and the classical confusion noise. For a discussion of transient detection in the visibility domain, see \citet{trott2011}. 

The problem of optimal detection of weak transient signals in an ambient noise has been treated with great care in the context of engineering applications (e.g. radars, communications, see for example~\citealt{helmstrom-1968}) as well as astronomical observations (e.g. gravitational waves, see~\citealt{Allen:2005fk} and references therein). The general approach to this problem is to define the appropriate statistical measure, the detection statistic $\rho(x)$ which is optimized using the Neyman--Pearson criteria, i.e. maximizing the probability of signal detection at a fixed probability of false alarm~\citep{neyman-1933}. The optimization leads to the quantity known as the likelihood ratio or the Bayes factor (for a derivation in the context of gravitational wave searches, see~\citealt{PhysRevD.85.122008}). In practice this quantity is usually estimated using the maximum likelihood approximation, in which marginalization over unknown parameters, such as the time of arrival and the amplitude of the signal, is replaced by maximization. If the signal has a known form, this approach results in the well-known matched filter technique~\citep{helmstrom-1968}. While in astronomical context the precise form of the signal is rarely known, the matched filter technique is still very useful, allowing one to increase the sensitivity of the search by narrowing down the space of admissible signals. In the case of radio observations, the standard matched filter must be modified to account for the classical confusion noise. 

The search for radio transients can be thought of as a two-stage process. At the first stage, the dirty snapshot images taken by the telescope and corrected for sky motion are stacked together in an image time-series. For every pixel (representing a synthesized beam) in the image, the time-series of the measured flux $x(t)$ is generated. The pixels could be defined on the grid in $(\mathrm{RA}, \mathrm{Dec})$ sampling the observed sky.  At the second stage, the measured flux time-series is searched for transient signals using the afterglow lightcurves as templates in a matched filter. In the blind survey, when the position of the transient is not known, each synthesized beam must be searched for the presence of a transient. The matched filter detection statistic for radio transients, derived in Appendix \ref{appendix_match_filter}, is maximized over the unknown signal amplitude and the classical confusion noise. It is given by      
\begin{equation}
\label{match_filter_def}
\rho(x) = \frac{(x,f-\langle f \rangle)}{\sqrt{(f-\langle f \rangle,f-\langle f \rangle)}}\,,
\end{equation}
where $x(t) \equiv \{x_1, x_2, \cdots x_N\}$ collectively denotes the time-series of the measured flux at times (snapshots) $\{t_1, t_2 \cdots t_N\}$, $f(t) \equiv \{f_1, f_2, \cdots f_N\}$ denotes the flux time-series as predicted by the template lightcurve with no noise added, $\langle f \rangle = \sum_{i = 0}^{i = N} f_i / N$ is the average signal flux over the whole observation, the inner product $(x,f-\langle f \rangle) =  \sum_{i = 0}^{i = N} x_i(f_i - \langle f \rangle)/\sigma_{\mathrm{th}}^2$ cross-correlates the data with the template weighted by the thermal noise, and $(f-\langle f \rangle,f-\langle f \rangle) = \sum_{i = 0}^{i = N} (f_i - \langle f \rangle)^2/\sigma_{\mathrm{th}}^2$ is the square of the norm of the template, which in principle can be set to unity by adjusting the overall amplitude of the template $f_i \to f_i/\sqrt{(f-\langle f \rangle,f-\langle f \rangle)}$. 

There are a few things worth noting about \Cref{match_filter_def}. The matched filter detection statistic $\rho(x)$ defines a linear filter for the data $x(t)$. The template $f(t)$ is defined across the whole observational period $\tau \ge t_{\sint}$ that typically consists of many snapshots and exceeds the transient duration. $f(t)$ is set to null at the time samples during which the transient is `off', and the corresponding data snapshots provide the reference images. The average template flux $\langle f \rangle$ is averaged over the whole observation and, as a result, goes to zero with increasing observation time as $1/\tau$. The appearance of $\langle f \rangle$ in \Cref{match_filter_def} owes itself to the presence of the classical confusion noise. $\rho(x)$ both estimates and `subtracts' it. Estimation and subtraction of the classical confusion noise improves with the accumulation of reference images and becomes precise as $\tau \to \infty$. When $\langle f \rangle \to 0$, we recover the standard expression for a matched filter in the presence of Gaussian noise, which means that the constant classical confusion noise is exactly `subtracted' out. The procedure of `subtracting' can be made explicit by realizing that $\rho(x)$ can be rewritten as $\rho(x) = (x-\langle x \rangle ,f)/\sqrt{(f-\langle f \rangle,f-\langle f \rangle)}$. In this way, it is clear that the average measured flux in the images is subtracted from the data before the data are cross-correlated with $f(t)$. 

Using the fact that $\rho(x)$ is linear in $x(t)$, it is easy to determine its statistical properties (see Appendix \ref{appendix_match_filter} for details). Namely, given that the underlying random noise in the data is Gaussian with variance $\sigma_{\mathrm{th}}$, the probability distribution for $\rho(x)$ in the case of the data containing only noise, $p(\rho \given 0)$, is also Gaussian with mean $\mu_0 = 0$ and variance $\sigma_0 =  \sqrt{(f-\langle f \rangle,f-\langle f \rangle)}$. If the data contain the signal $Af(t)$, the probability distribution for the detection statistic $p(\rho \given 1)$ is still Gaussian with the same variance $\sigma_1 =  \sqrt{(f-\langle f \rangle,f-\langle f \rangle)}$ and the mean shifted in proportion to the signal amplitude $\mu_1 = A(f-\langle f \rangle,f-\langle f \rangle)$. Note that the shift is proportional to the signal amplitude $A$. 

The parameters describing the transient signal, such as the overall amplitude $A$ and the time of arrival $t_{\mathrm{a}}$ are unknown. In particular, $t_{\mathrm{a}}$ must be determined by shifting the template $f(t)$ in time, evaluating $\rho(x, t)$ as a function of time, and searching for the maximum in the resulting time-series $\rho(x) = \max_{t} \rho(x, t)$. Following this procedure, one can estimate $t_{\mathrm{a}}$ as $t_{\mathrm{a}} = \arg \max_t \rho(x, t)$. The detection statistic given by \Cref{match_filter_def} is already maximized over $A$, which can be computed from $\rho(x)$ using \Cref{best_amplitude}. In the case of afterglows, $A$ and $t_{\mathrm{a}}$ are not the only parameters characterizing them. Their brightness and duration depend strongly on $E_{\jet}$, $\theta_{\obs}$, and $n$. As these parameters are unknown as well, the usual strategy is to prepare a bank of templates $f(E_{\jet},\theta_{\obs},n)$ sampling this parameter space and evaluate the matched filter for every template $\rho(x,E_{\jet},\theta_{\obs},n)$. As in the case of an unknown $t_{\mathrm{a}}$, one can maximize over these extra parameters by estimating them from the data. The highest value of the matched filter $\rho(x) = \max_{(E_{\jet},\theta_{\obs},n,t)} \rho(x,E_{\jet},\theta_{\obs},n,t)$ constitutes the detection statistic. In the search, $\rho(x)$, computed for every pixel, should be compared with the distribution of $\rho(x)$ one expects to find if the data contain no transient signal, $p(\rho \given 0)$. Using this distribution and accounting for the number of independent trials (coming, for example, from searching many pixels with a bank of templates), one can compute the probability of false alarm for every candidate in the search. For a signal to be detected with a high degree of confidence, the false alarm probability should be lower than the probability to find a $5\sigma$ deviation for a normal random variable $p(\rho \given 0) \le [1 - \mathrm{erf}(5/\sqrt{2})] \approx 5.7 \times 10^{-7}$. 

The full implementation of the detection procedure outlined above in a real-life search requires a separate investigation. However, for the purpose of establishing a realistic criteria for afterglow detectability, it is sufficient to consider a simplified version of the analysis which, nevertheless, captures the key aspects of the detection of radio transients. The main simplification comes from characterizing the afterglow lightcurves only by their peak flux $f_{\mathrm{peak}}$ and duration $t_{\mathrm{dur}}$. The corresponding template $f(t)$ is the top-hat function with $f_i = f_{\sref}$ when the transient is on and zero otherwise. The template reference flux $f_{\sref}$ is $f_{\mathrm{peak}}$ of a source at the reference distance $d_{\sref}$. Neglecting the effects of maximization over time and the template bank and using the statistical properties of $\rho(x)$, we find (see Appendix \ref{appendix_match_filter} for details) that the variance of $p(\rho \given 0)$ is 
\begin{equation}
\label{sigma_0}
\sigma_0 = \frac{f_{\sref}}{\sigma_{\thermal}}\sqrt{\frac{N - 1}{N}}\,,
\end{equation}
where $N = \tau/t_{\mathrm{dur}}$ is the duration of the observation in units of the transient duration. For robust signal detection, we impose the condition for the mean of $p(\rho \given 1)$ to be $7\sigma_0$ away from the mean of $p(\rho \given 0)$ . This is equivalent to achieving $97\%$ or higher efficiency in detecting signals at the $5\sigma$ threshold on the false alarm probability (see \Cref{fig:det_stats_dist} for illustration and Appendix \ref{appendix_match_filter} for explanation). The condition is satisfied for the signals with peak fluxes
\begin{equation}
\label{flux_threshold_def}
f_{\peak} \ge f^{\ast}=7\sqrt{\frac{N}{N-1}}\, \sigma_{\thermal}\,.
\end{equation}
For observations that are much longer than the transient duration, $N \to \infty$, the threshold flux approaches what would be achievable in the absence of the classical confusion noise, $f^{\ast} \to 7 \sigma_{\thermal}$. The afterglows tend to be long-lasting, so in practice the observations will span at best a few transient durations. We choose $N = 2$, allowing for the reference image to be as long as the transient itself. In addition, we impose the upper limit of 3\,yr on $t_{\dur}$ independent of $f_{\peak}$. In the absence of archival radio data that can serve as references, it seems impractical to attempt detecting afterglows significantly longer than that. Combining the two thresholds, we impose the following detectability criteria:
\begin{align}
f_{\peak}  &\ge 7\sqrt{2}\, \sigma_{\thermal} \,, \label{eq:f_thres} \\ 
t_{\mathrm{dur}} & \le 3\text{yr} \,. \label{eq:criteria}
\end{align}

\section{Rate Estimation}\label{rates}

Having defined a flux threshold and a duration threshold to characterize the detectability of afterglows with radio instruments in Section~\ref{detectability}, we estimate the number of \ac{SGRB} afterglows that we expect an instrument to detect given the intrinsic rate of these events as well as the sensitivity, field of view, and survey strategy of the instrument. 

The association between \ac{SGRB}s and \ac{BNS} coalescence is promising but far from conclusive. None the less, the intrinsic rate of \ac{SGRB}s as derived from \ac{SGRB} observations is consistent with the rates of \ac{BNS} coalescence as derived from binary pulsar observations and population synthesis. Hence, we assume the rate of \ac{SGRB}s is equal to that of \ac{BNS} coalescence and use the predicted rates from \citet{abadie2010}, who derive pessimistic, realistic, optimistic, and upper limit estimates. Uncertainties in these estimates, spanning three orders of magnitude, dominate the uncertainties in our estimates of detection rates. Consequently, detections or non-detections from radio instruments may confirm or rule out the optimistic models of \ac{BNS} coalescence and decrease the uncertainty in the expected rates for \ac{GW} detectors. 

Given the intrinsic rate of \ac{SGRB} afterglows $R_{\cbc}$ (number per volume, per year), we calculate the rate of afterglow detections $R_{\det}$ (number per year) expected for a radio instrument by determining the volume that the instrument can observe. This volume depends on the sensitivity or flux threshold $f^{\thres}$ of the instrument and the sky area $\Omega_{S}$ covered by the survey. 

$f^{\thres}$ sets the maximum luminosity distance $d_{\mathrm{L}}$ to which the instrument can observe a source with a fixed luminosity. We convert $d_{\mathrm{L}}$ to a horizon distance $d_{\mathrm{H}}$ that we define to be the comoving distance corresponding to $d_{\mathrm{L}}$ at redshift $z'$, both of which are unknown and need to be computed:
\begin{subequations} \label{eq:dh}
\begin{align}
&d_{\mathrm{H}} = \frac{d_{\mathrm{L}}(z')}{1+z'} \label{eq:dh_def}\\ 
\mbox{where} \ \ &\frac{d_{\mathrm{L}}^{2}(z')}{1+z'} = d^{2}_{\sref} \left( \frac{f_{\sref}}{f^{\thres}} \right) \label{eq:dl_def}\,.
\end{align}
\end{subequations}
$d_{\sref}$ is the reference distance at which the afterglow lightcurves are generated, and $f_{\sref}$ is the peak flux of the afterglow at $d_{\sref}$. The factor of $(1+z')$ in \Cref{eq:dl_def} is the $k$-correction term \citep{hogg1999}. We assume the \planck~2013 cosmology \citep{planck2013}. 

\begin{figure}
  \centering
  \includegraphics[width=0.5\textwidth]{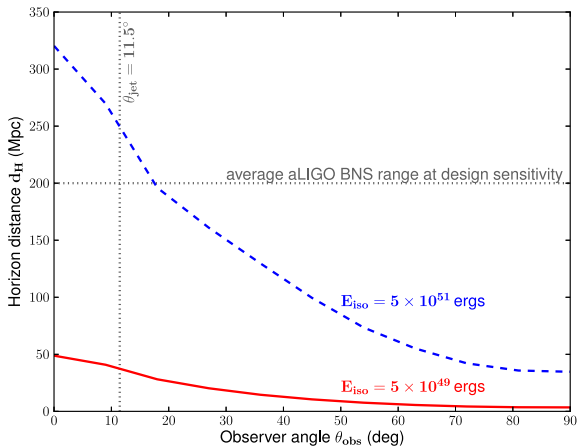}
  \caption{MWA horizon distances for afterglows with different $E_{\iso}$ but in the same environment ($n = 10^{-3}$\,cm$^{-3}$). $\nu_{\obs} = 150$\,MHz and $f^{\thres} = 0.8$\,mJy. The vertical dotted line marks the jet opening angle. Afterglows that are more energetic or more on-axis are brighter and therefore detectable to larger distances. For comparison, the average range of \ac{BNS} coalescence for aLIGO at design sensitivity is 200\,Mpc, as illustrated by the horizontal dotted line \citep{ligo2013}.}
  \label{fig:dh_energy}
\end{figure}

\begin{figure}
  \centering
  \includegraphics[width=0.5\textwidth]{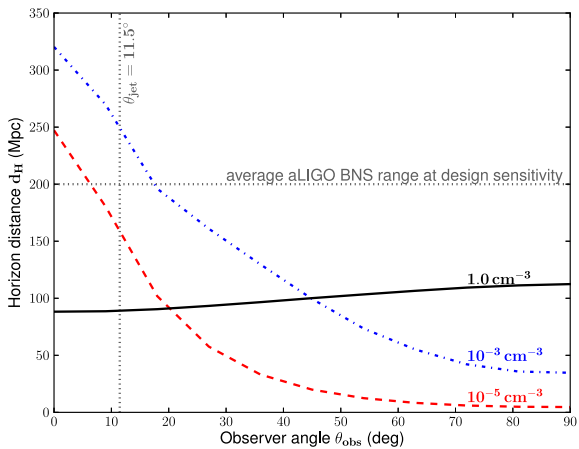}
  \caption{MWA horizon distances for afterglows in different environments but with the same $E_{\iso} = 5 \times 10^{51}$\,ergs. $\nu_{\obs} = 150$\,MHz and $f^{\thres} = 0.8$\,mJy. The vertical dotted line marks the jet opening angle. Afterglows become brighter as $n$ increases until synchrotron self-absorption becomes dominant. The $d_{\mathrm{H}}$ curve for $n = 1$\,cm$^{-3}$ (black solid line) is almost independent of $\theta_{\obs}$ because emission is isotropic by the time the system becomes optically thin. Contribution from the counter-jet makes the off-axis afterglows slightly brighter than the on-axis ones for $n = 1$\,cm$^{-3}$.}
  \label{fig:dh_density}
\end{figure}

By definition, the same instrument will have a range of $d_{\mathrm{H}}$ corresponding to different lightcurves with different $f_{\sref}$. To calculate $d_{\mathrm{H}}$ for each instrument, we substitute the corresponding $f^{\thres}$ into \Cref{eq:dh} and numerically solve for $z'$ (hence $d_{\mathrm{L}}$ and $d_{\mathrm{H}}$) for the lightcurves we generate in Section~\ref{lightcurves}. \Cref{fig:dh_energy,fig:dh_density} show example horizon distances for the MWA. The shapes of the $d_{\mathrm{H}}$ curves trace the variations of $f_{\sref}$ as a function of $\theta_{\obs}$ and $E_{\iso}$ while the normalization is set by $f^{\thres}$ of the instrument. In other words, another instrument operating at the same frequency but with a different sensitivity will have $d_{\mathrm{H}}$ curves of approximately the same shape (up to cosmological corrections) but a different amplitude. 

Since $d_{\mathrm{H}}$ depends on $\theta_{\obs}$ and $E_{\iso}$, we integrate over $\theta_{\obs}$ and $E_{\iso}$ when we calculate $R_{\det}$:  
\begin{equation}\label{eq:detection_rate}
\begin{split}
R_{\det} = &\left[ \frac{R_{\cbc}}{4\pi (E_{2} - E_{1})} \right] \times \\
& \int_{E_{1}}^{E_{2}} \int_{4\pi} \left[\Omega_{S}\frac{d_{\mathrm{H}}^{3}(\theta_{\obs}, E_{\iso})}{3}\right] \diff \Omega_{\obs} \diff E_{\iso}\,.
\end{split}
\end{equation}
This equation assumes that the bursts are uniformly distributed in energy ($5 \times 10^{49} \leq E_{\iso} \le 5 \times 10^{51}$\,ergs) and jet orientation ($-1 \leq \cos \theta_{\obs} \leq 1$). It also treats each $n$ separately, where $n = 10^{-5}$\,cm$^{-3}$ represents the intergalactic medium (outside the host galaxy) and $n = 1$\,cm$^{-3}$ represents the interstellar medium (inside the host galaxy). If \ac{SGRB}s occur equally likely in the different environments, $R_{\det}$ would be the average of the separate values. To integrate over $E_{\iso}$, we use the analytical energy-flux scaling relation for $f_{\sref}$ derived by \citet{vaneerten2012} (see also Section~\ref{lightcurves}) when we calculate $d_{\mathrm{H}}$. If $d_{\mathrm{H}}$ is independent of $\theta_{\obs}$ and $E_{\iso}$, \Cref{eq:detection_rate} reduces to $R_{\det} = R_{\cbc} (\Omega_{S} d_{\mathrm{H}}^{3}/ 3)$. 

\begin{figure}
  \centering
  \includegraphics[width=0.5\textwidth]{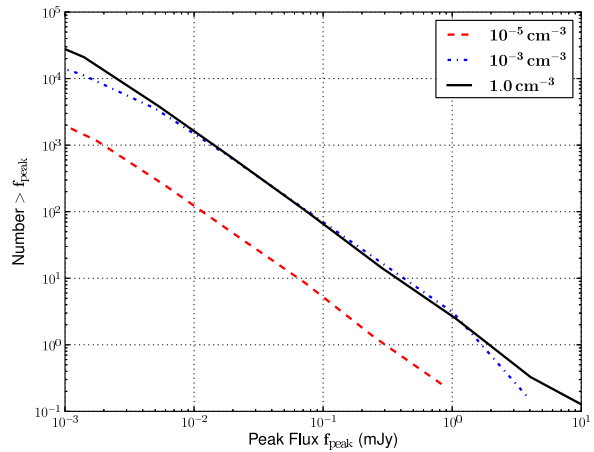}
  \caption{Cumulative distributions of $f_{\peak}$ for the simulated afterglow lightcurves at 150\,MHz. This is similar to \Cref{fig:cumulative_flux_intrinsic} but plotted for afterglows with $t_{\dur} \leq 3$\,yr. While certain afterglows could be bright enough to be detected as sources (see \Cref{fig:cumulative_flux_intrinsic}), they may last longer than the survey and therefore not be detected as transients.}
  \label{fig:cumulative_flux_cut}
\end{figure}

\begin{figure}
  \centering
  \includegraphics[width=0.5\textwidth]{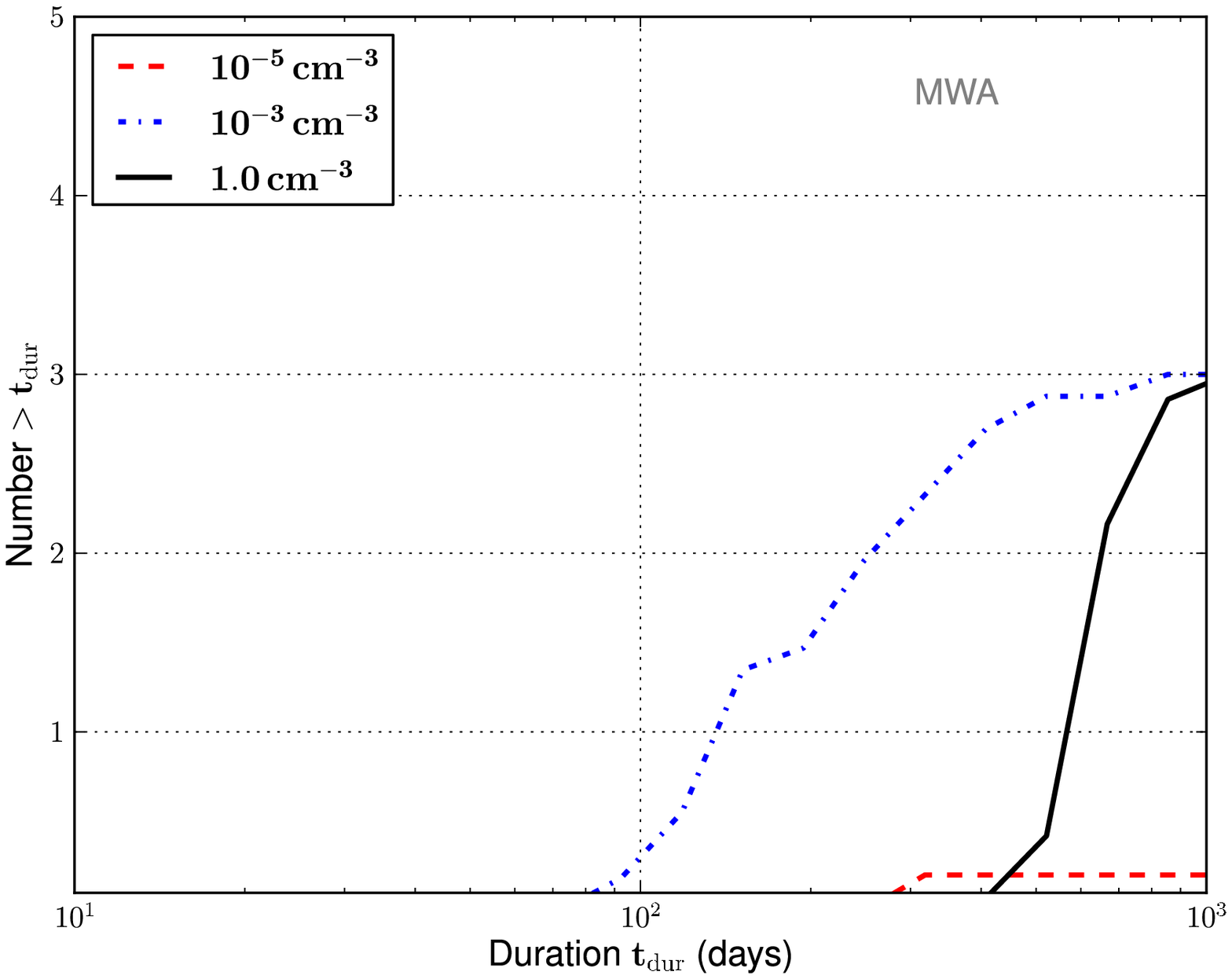}
  \includegraphics[width=0.5\textwidth]{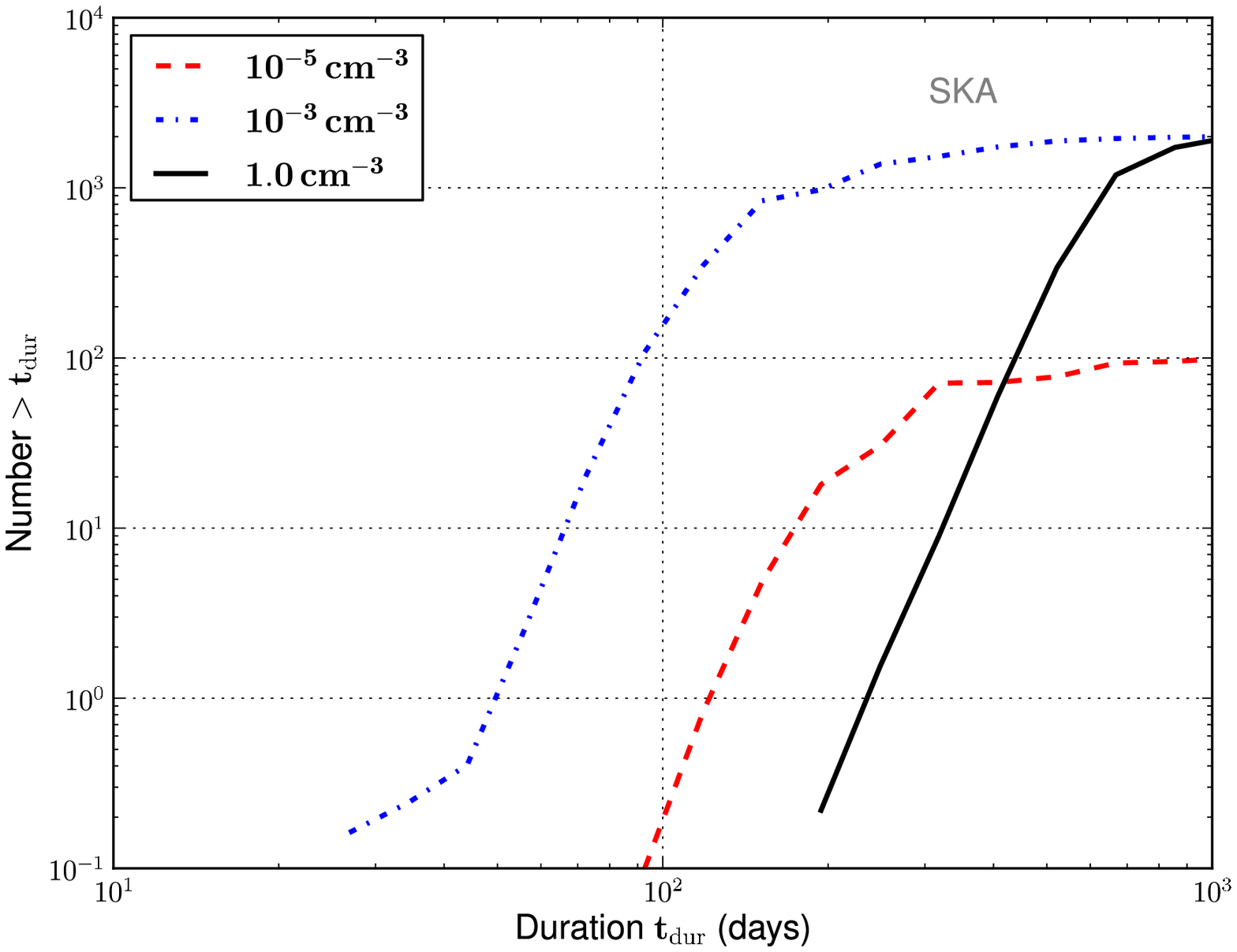}
  \caption{Cumulative distributions of $t_{\dur}$ for the simulated afterglow lightcurves at 150\,MHz. This is plotted for afterglows with $t_{\dur} \leq 3$\,years and $f_{\peak} \ge f^{\thres}$ for MWA ($f^{\thres} = 0.8$\,mJy; top) and SKA ($f^{\thres} = 0.008$\,mJy; bottom). As most of these afterglows last $\sim 1$\,yr, a survey to detect these events should revisit the same area of the sky on a similar time-scale. At higher frequencies, the durations are shorter ($\ga 3$\,months).}
  \label{fig:cumulative_duration}
\end{figure}

During the calculation of $R_{\det}$, we impose a cut on afterglow duration according to \Cref{eq:criteria}. An afterglow that lasts longer than the survey or the availability of archival data will not be detected as a transient event even if it is bright. \Cref{fig:cumulative_flux_cut} shows the cumulative distribution of peak fluxes for afterglow lightcurves with the constraint that $t_{\dur} \leq 3$\,yr. This particular choice of $t_{\dur}$ manages to capture the majority of the detectable population without requiring a survey to last an impractical length of time. At 150\,MHz, most of these afterglows last $\sim 1$\,yr, as evident from \Cref{fig:cumulative_duration}, suggesting that a survey should cover a time range that is at least as long. At higher frequencies, the durations are shorter ($\ga 3$\,months).

We now present results for $R_{\det}$ for three types of observations: blind surveys, \ac{SGRB} follow-up observations, and \ac{GW} candidate follow-up observations.

\subsection{Blind Survey}\label{blind_survey}

There are two possible strategies for blind surveys: `narrow and deep' or `shallow and wide', where we require the total time allocated for the survey to last much longer than the time needed to reach a good thermal sensitivity in a snapshot image. \Cref{eq:dh,eq:f_thres} show that $d_{\mathrm{H}} \propto \sigma_{\thermal}^{-1/2}$. Combining this relation with $R_{\det} \propto \Omega_{S}d_{\mathrm{H}}^{3}$ from \Cref{eq:detection_rate} and $\sigma_{\thermal} \propto t_{\sint}^{-1/2}$ from \Cref{eq:thermal_noise}, we get the following dependence: 
\begin{equation}\label{eq:rate_scaling}
R_{\det} \propto \Omega_{S}t_{\sint}^{3/4}\,.
\end{equation}
This shows that $R_{\det}$ increases faster with $\Omega_{S}$ than it does with $t_{\sint}$, arguing in favor of a `shallow and wide' approach. In other words, for a fixed survey length $t_{S} \gg t_{\sint}$, the survey should maximize its sky coverage over the time $t_{S}$ rather than performing a deep exposure on a small patch of the sky if the goal of the survey is to increase $R_{\det}$. 

An all-sky survey covers the maximal area that any instrument can observe. The amount of accessible sky varies with the location on Earth while the Galactic plane obscures extragalactic observations. To account for this effect, we choose $\Omega_{S} = 2\pi$ for every radio instrument that we consider in our analysis. As each instrument has an instantaneous field of view $\Omega_{\fov} \leq \Omega_{S}$, it needs $N_{p}$ separate pointings, each with integration time $t_{\sint}$, to cover the total survey area $\Omega_{S} = N_{p}\Omega_{\fov}$. We consider a survey length of $t_{S} = 1$\,yr, assuming $100\%$ duty cycle. Thus, $\sigma_{\thermal}$ (hence $f^{\thres}$) is set by $t_{\sint} = t_{S} / 2N_{p}$, where the factor of 2 comes from our requirement that the instrument observe the same patch of the sky twice to detect a source as a transient. Results for selected instruments are listed in \Cref{tb:blind_survey}.
\begin{table*}
  \centering
  \begin{minipage}{165mm}
  \caption{Expected rates of afterglow detection for radio instruments conducting a blind survey, grouped according to the circumburst environments and rounded to two significant digits. $\nu_{\obs}$ is the observer frequency at which we generate the afterglow lightcurves and $f^{\thres}$ is the flux threshold of each instrument. The rates only include afterglows with $f_{\peak} \ge f^{\thres}$ and $t_{\dur} \leq 3$\,yr (see Section~\ref{detectability} for more details). The sky coverage for all radio instruments is $2\pi$ (an all-sky survey crudely corrected for Earth location and Galactic plane exclusion). The LIGO detectors have $4\pi$ coverage of the sky. We use the \ac{BNS} coalescence rates from \citet{abadie2010} to derive our estimates and include the estimates for Initial LIGO as reference. For Advanced LIGO, we compute the expected rates of detection directly from $R_{\cbc}$ as $4 \pi {(D^{\mathrm{GW}}_{\mathrm{horizon}}/2.26)}^3/3 \times R_{\cbc} \approx 33\times10^{6}$\,Mpc$^{3} \times R_{\cbc}$, where $D^{\mathrm{GW}}_{\mathrm{horizon}} = 450$\,Mpc is the aLIGO horizon distance for an optimally-located and oriented binary while the factor of 2.26 accounts for the averaging done over the sky and all possible orientations. Uncertainties in the \ac{BNS} coalescence rates dominate uncertainties in our estimates of the afterglow detection rates.}
  \begin{tabular}{lccccccc} 
    \hline
    & $\nu_{\obs}$ (MHz)  & $f^{\thres}$ (mJy) & Pessimistic & Realistic & Optimistic & Upper Limit &  \\ 
    \hline 
    BNS coalescence  & - & -  & 0.01 & 1 & 10 & 50 & Mpc$^{-3}$\,Myr$^{-1}$ \\ 
    Initial LIGO & - & - & $2\times 10^{-4}$ & 0.02 & 0.2 & 0.6 & yr$^{-1}$ \\
    Advanced LIGO & - & - & 0.33 & 33 & 333 & 1665 & yr$^{-1}$ \\ 
    \hline
    \multicolumn{8}{c}{$n = 10^{-5}$\,cm$^{-3}$ (outside host galaxy)} \\ 
    \hline 
    LWA1 & 60 & 23.3 & $1 \times 10^{-5}$ & 0.001 & 0.01 & 0.05 & yr$^{-1}$ \\
    LOFAR Low & 60 & 6.42 & $7 \times 10^{-5}$ & 0.007 & 0.07 & 0.3 & yr$^{-1}$ \\
    LOFAR High & 150 & 0.81 & 0.002 & 0.2 & 2 & 9 & yr$^{-1}$ \\
    MWA & 150 & 0.80 & 0.002 & 0.2 & 2 & 9 & yr$^{-1}$ \\
    CHIME Pathfinder & 600 & 0.176 & 0.02 & 2 & 20 & 100 & yr$^{-1}$ \\
    CHIME Full & 600 & 0.026 & 0.3 & 30 & 350 & 2000 & yr$^{-1}$ \\
    ASKAP & 1430 & 0.114 & 0.03 & 3 & 30 & 200 & yr$^{-1}$ \\
    HERA & 150 & 0.012 & 0.09 & 9 & 90 & 500 & yr$^{-1}$ \\ 
    SKA1 Low & 150 & 0.008 & 1 & 100 & 1000 & 6000 & yr$^{-1}$ \\ 
    \hline
    \multicolumn{8}{c}{$n = 10^{-3}$\,cm$^{-3}$} \\ 
    \hline
    LWA1 & 60 & 23.3 & $7 \times 10^{-5}$ & 0.007 & 0.07 & 0.4 & yr$^{-1}$ \\
    LOFAR Low & 60 & 6.42 & $5 \times 10^{-4}$ & 0.05 & 0.5 & 2 & yr$^{-1}$ \\
    LOFAR High & 150 & 0.81 & 0.03 & 3 & 30 & 150 & yr$^{-1}$ \\
    MWA & 150 & 0.80 & 0.03 & 3 & 30 & 150 & yr$^{-1}$ \\
    CHIME Pathfinder & 600 & 0.176 & 0.5 & 50 & 500 & 2500 & yr$^{-1}$ \\
    CHIME Full & 600 & 0.026 & 6 & 650 & 6000 & $3 \times 10^{4}$ & yr$^{-1}$ \\
    ASKAP & 1430 & 0.114 & 0.8 & 80 & 800 & 4000 & yr$^{-1}$ \\
    HERA & 150 & 0.012 & 1 & 100 & 1000 & 6500 & yr$^{-1}$ \\ 
    SKA1 Low & 150 & 0.008 & 20 & 2000 & $2 \times 10^{4}$ & $8 \times 10^{4}$ & yr$^{-1}$ \\ 
    \hline
    \multicolumn{8}{c}{$n = 1.0$\,cm$^{-3}$ (inside host galaxy)} \\ 
    \hline
    LWA1 & 60 & 23.3 & $6 \times 10^{-6}$ & $6 \times 10^{-4}$ & 0.006 & 0.03 & yr$^{-1}$ \\
    LOFAR Low & 60 & 6.42 & $4 \times 10^{-5}$ & 0.004 & 0.04 & 0.2 & yr$^{-1}$ \\
    LOFAR High & 150 & 0.81 & 0.03 & 3 & 30 & 150 & yr$^{-1}$ \\
    MWA & 150 & 0.80 & 0.03 & 3 & 30 & 150 & yr$^{-1}$ \\
    CHIME Pathfinder & 600 & 0.176 & 2 & 200 & 2000 & $1 \times 10^{4}$ & yr$^{-1}$ \\
    CHIME Full & 600 & 0.026 & 30 & 3000 & $3 \times 10^{4}$ & $1 \times 10^{5}$ & yr$^{-1}$ \\
    ASKAP & 1430 & 0.114 & 8 & 800 & 8000 & $4 \times 10^{4}$ & yr$^{-1}$ \\
    HERA & 150 & 0.012 & 2 & 200 & 2000 & 8000 & yr$^{-1}$ \\ 
    SKA1 Low & 150 & 0.008 & 20 & 2000 & $2 \times 10^{4}$ & $1 \times 10^{5}$ & yr$^{-1}$ \\ 
    \hline
  \end{tabular}
  \label{tb:blind_survey}
  \end{minipage}
\end{table*}
These results are computed for $\Omega_{S} = 2\pi$ sky area and depend on $f^{\thres}$ (i.e. $t_{\sint}$), so one should use the dependence in \Cref{eq:rate_scaling} to obtain the rates for a different sky coverage or flux sensitivity. 

Radio instruments currently operating at very low frequencies ($<80$\,MHz) will not be sensitive to afterglows because of strong synchrotron self-absorption and modest instrumental sensitivities. The other instruments, however, may be able to constrain optimistic \ac{BNS} rate predictions before the advanced \ac{GW} detectors begin operating. As synchrotron emission is stronger at higher densities, all radio instruments will be more sensitive to events occuring in a dense medium. Future instruments will be able to detect many afterglows even on a relatively short time-scale ($<3$\,yr). Our results are optimistic as we assume $100\%$ duty cycle of these instruments, but they give an order of magnitude estimate of what surveys with these instruments might see. Actual numbers will depend on the survey details and achieved sensitivities. Uncertainties in the sidelobe confusion noise, the primary beam, or calibration will lower the sensitivity of the instrument. 

\subsection{Gamma-Ray Burst Follow-up}

\ac{SGRB} follow-up observations will be sensitive only to the population of on-axis afterglows. To calculate $R_{\det}$ for on-axis afterglows, we use the constraint $0 \leq \theta_{\obs} \leq \theta_{\jet}$ instead of $0 \leq \theta_{\obs} \leq \pi/2$ when we integrate \Cref{eq:detection_rate}. $\theta_{\jet} = 11.5\degr$ for our simulated lightcurves (\Cref{tb:simulation_parameters}), consistent with observed values of $\theta_{\jet}$ although the uncertainty is quite large (see \citealt{berger2013_review} and references therein). A larger $\theta_{\jet}$ implies a larger fraction of on-axis afterglows. 

Instead of estimating the number of SGRBs that the radio instruments can detect given a SGRB trigger from a $\gamma$-ray telescope such as \textit{Swift} or \textit{Fermi}, we present estimates of the fraction of on-axis bursts present in the blind surveys (\Cref{tb:grb_followup}). These are bursts that, in principle, could have $\gamma$-ray counterparts and may trigger a $\gamma$-ray telescope. While we account for the instantaneous fields of view of \textit{Swift} and \textit{Fermi} \citep{siellez2013}, we do not consider other factors such as the systematic uncertainties of $\gamma$-ray detectors and selection effects that could lower the fraction of bursts detectable by $\gamma$-ray telescopes. The fraction of on-axis afterglows increases with decreasing circumburst density because as the synchrotron emission becomes weaker, the radio instruments become less sensitive to off-axis afterglows and detect only the population that is more on-axis. 

\begin{table}
  \centering
  \caption{Fraction of on-axis radio afterglows in a blind survey. These could be accompanied by $\gamma$-ray counterparts and trigger a $\gamma$-ray telescope. We account for the fields of view of \textit{Swift} (1.4\,sr) and \textit{Fermi} (9.5\,sr), which cover 0.867 of the entire sky without overlapping regions \citep{siellez2013}, but not any systematics specific to $\gamma$-ray detection.}
  \begin{tabular}{lccc} 
    \hline
    Instrument & $10^{-5}$\,cm$^{-3}$ & $10^{-3}$\,cm$^{-3}$ & $1.0$\,cm$^{-3}$ \\ 
    \hline
    LWA1 & 0.58 & 0.22 & 0.002 \\
    LOFAR Low & 0.58 & 0.22 & 0.002 \\
    LOFAR High & 0.60 & 0.37 & 0.01 \\
    MWA & 0.60 & 0.37 & 0.01 \\
    CHIME Path & 0.68 & 0.49 & 0.04 \\
    CHIME & 0.66 & 0.46 & 0.04 \\
    ASKAP & 0.72 & 0.55 & 0.08 \\
    HERA & 0.57 & 0.32 & 0.01 \\ 
    SKA1 Low & 0.57 & 0.31 & 0.01 \\ 
    \hline
  \end{tabular}
  \label{tb:grb_followup}
\end{table}

As some afterglows detectable in a blind radio survey are on-axis and could have $\gamma$-ray counterparts, coincident detections at different wavelengths could increase the significance of weak signals. While \textit{Swift} detections are well-localized, \textit{Fermi} detections often have large localization uncertainties (10--100\,deg$^{2}$). Consequently, many \textit{Fermi} detections do not have follow-up observations at other wavelengths (cf. \citealt{singer2013}). Current widefield radio instruments could develop \ac{SGRB} follow-up strategies with \textit{Fermi} to guide future \ac{GW} follow-up strategies.

\subsection{Gravitational Wave Candidate Follow-up}

\ac{GW} candidates lie within the horizon distance of the \ac{GW} detector. If this distance is less than $d_{\mathrm{H}}$ of the radio instrument performing the follow-up observation, all \ac{GW} events will be detectable, but this is more often not the case. 

To estimate the expected number of afterglow detections given the detection of a \ac{GW} event by aLIGO or a similar ground-based detector, we assume that a \ac{GW} signal from an optimally-located and oriented binary can be detected from up to 450\,Mpc, the designed sensitivity of aLIGO \citep{abadie2010}. In general, the distance at which the \ac{GW} signal from a \ac{CBC} is detectable depends on the location of the binary on the sky, the inclination, and the polarization angle. The horizon distance for the radio signal depends on the jet energy, the jet orientation, and the circumburst density. While the current models of jet formation predict that the SGRB jet is likely to be aligned with the inclination of the binary (binary disc is face-on when the jet is on-axes; see \citealt{narayan1992} and \citealt{kochanek1993}), observationally the question is far from being settled. Consequently, we explore two distinct cases: (i) the jet is aligned with the binary inclination, (ii) the jet orientation and the binary inclination are completely uncorrelated. For both scenarios, we compute the fraction of detected \ac{GW} signals that are also detectable with a radio telescope, averaging over all intrinsic and extrinsic parameters of the binary and the SGRB. See Appendix \ref{appendix_ligo_followup} for the details of these calculations.

The calculation of $R_{\det}$ for \ac{GW} follow-up is otherwise similar to that of a blind survey. Unlike a blind survey, $t_{S}$ is divided by the expected number of aLIGO events or the number of pointings needed to cover the entire sky, whichever is smaller. For each event, we use the expected aLIGO localization error as $\Omega_{S}$, choosing $100$\,deg$^{2}$ as our value \citep{ligo2013}. Most of the radio instruments, however, cover this error box in one pointing. If aLIGO detects many events such that the sky surface density is high, the observation strategy for a radio instrument then becomes indistinguishable from that of a blind survey. \Cref{tb:ligo_followup} presents estimates of the fraction of aLIGO events that various radio instruments could detect. Following up all the aLIGO events requires at least two telescopes, one in the northern and one in the southern hemisphere. Future radio instruments that might be operating at the same time as aLIGO will likely be able to detect most or all of the aLIGO events, except when bursts occur at the lowest densities. 

\begin{table}
  \centering
  \caption{Average fraction of \ac{BNS} events detectable by both aLIGO and radio follow-up observations. The fraction is normalized to $2\pi$ sky area accessible to a radio telescope. Following up on all of the aLIGO events requires telescopes in the northern and the southern hemispheres. Results for two cases are listed $ \star / \star$: \textit{the jet is aligned with the binary inclination}~/~\textit{the jet is uncorrelated with the binary inclination}. HERA is suboptimal for aLIGO follow-up observations because it is a drift-scan telescope and cannot point to cover the whole sky, unlike CHIME, which sees the whole northern hemisphere. The horizon distance that we use for aLIGO is 450\,Mpc, the value for an optimally-located and oriented \ac{BNS} system for aLIGO at design sensitivity. These results are optimistic as we assume that the instruments dedicate $100\%$ of their time to the follow-up observations.}
  \begin{tabular}{lccc} 
    \hline
    Instrument & $10^{-5}$\,cm$^{-3}$ & $10^{-3}$\,cm$^{-3}$ & $1.0$\,cm$^{-3}$ \\ 
    \hline
    LWA1 & 0.001~/~0.005 & 0.01~/~0.03 & 0.01~/~0.01 \\
    LOFAR Low & 0.002~/~0.01 & 0.01~/~0.05 & 0.01~/~0.01 \\
    LOFAR High & 0.02~/~0.05 & 0.19~/~0.27 & 0.58~/~0.58 \\
    MWA & 0.02~/~0.05 & 0.18~/~0.27 & 0.57~/~0.57 \\
    CHIME Path & 0.05~/~0.10 & 0.30~/~0.36 & 0.96~/~0.96 \\
    CHIME & 0.12~/~0.15 & 0.55~/~0.54 & 1.0~/~1.0 \\
    ASKAP & 0.10~/~0.12 & 0.43~/~0.45 & 1.0~/~1.0 \\
    HERA & 0.02~/~0.02 & 0.09~/~0.09 & 0.13~/~0.13 \\ 
    SKA1 Low & 0.16~/~0.17 & 0.68~/~0.67 & 1.0~/~1.0 \\ 
    \hline
  \end{tabular}
  \label{tb:ligo_followup}
\end{table}

\section{Discussion}\label{discussion}

\ac{EM} follow-up of \ac{GW} candidates is important to establish the astrophysical nature of these events and to advance the study of their progenitors. However, the next generation of \ac{GW} detectors will have large localization errors during the early days of their operation, presenting a challenge for \ac{EM} follow-up efforts. As shown in this paper, new widefield radio instruments will have the capability to perform follow-up observations of \ac{SGRB} afterglows, a possible \ac{EM} counterpart of \ac{GW} events. 

While previous studies of \ac{SGRB} radio afterglows argued that these events are too faint and long-lasting to be detectable by present and planned instruments (\citealt{nakar2011}, \citealt{metzger2012}, \citealt{kelley2013}), this paper showed a spread in the distributions of peak fluxes and durations of these afterglow lightcurves, within a plausible range of model parameters, generated from the numerical tool \textsc{boxfit} developed by \citealt{vaneerten2012_boxfit} (Section~\ref{lightcurves}). These distributions are consistent with previous estimates, showing that most afterglows are faint ($\la \umu$Jy) and long-lasting ($\ga$\,yr). However, there is a tail of bright ($\sim$\,mJy) and short ($\sim$\,yr) afterglows that could be detectable by current and future radio instruments. This tail, however, is sensitive to the model parameters, such as $E_{\iso}$ and $\theta_{\jet}$, many of which are currently uncertain. The results in \citet{kelley2013}, who consider \ac{SGRB} radio afterglows as triggers to \ac{GW} searches, are more pessimistic than ours because they explored the low energy and high density ends of the plausible afterglow parameter space. Future studies exploring the dependence of the properties of radio afterglows on the various model parameters, such as a wider range of $\theta_{\jet}$, are needed. These late-time afterglows could also be a factor of a few brighter than previously expected \citep{sironi2013}, but this effect is not included in our study. 

To characterize the detectability of these afterglows, we derived a criteria on peak flux and duration based on a simple matched filter technique for radio instruments in the presence of thermal noise and constant noise from background source confusion (Section~\ref{detectability}). The actual sensitivies of radio instruments will be limited by other sources of error, such as calibration errors, primary beam errors, sidelobe confusion, etc. These are specific to the analyses performed with each instrument and are beyond the scope of this paper. None the less, our criteria provides an order of magnitude estimate for \ac{SGRB} afterglows that could be detectable by these instruments. The actual rates measured by these instruments will be lower because of instrumental systematics. False positives from intrinsic variability of other sources, such as AGN variability, may decrease the significance of detected events but could be distinguished using counterparts at other wavelengths. 

Converting the detectability criteria into a horizon distance, we estimated the rates of \ac{SGRB} afterglow detection expected for various radio instruments performing three types of surveys: blind surveys, \ac{SGRB} follow-up observations, and \ac{GW} follow-up observations (Section~\ref{rates}). Given the context of \ac{EM} follow-up of \ac{GW} events, we assumed the intrinsic rate of the progenitors of \ac{SGRB}s to be equal to that of \ac{BNS} coalescence as summarized in \citet{abadie2010}. Uncertainties on the \ac{BNS} rates dominate uncertainties in our estimations. Blind all-sky surveys with current widefield radio instruments will be able to constrain certain optimistic predictions for \ac{BNS} coalescence before the advanced \ac{GW} detectors turn on. They will also be able to characterize the background radio transients for future follow-up observations as well as perform independent studies of afterglows. A large fraction of afterglows in these blind surveys will also be on-axis bursts, suggesting that many detectable radio afterglows could have $\gamma$-ray counterparts that could trigger $\gamma$-ray telescopes. Coincident detections at different wavelengths could increase the significance of weak signals. Furthermore, \textit{Fermi} detections of \ac{SGRB}s could have large localization errors not unlike those of aLIGO. Strategies on \ac{SGRB} follow-up observations with current radio instruments could thus guide strategies on \ac{GW} follow-up observations with future instruments that will likely have the ability to detect most or all of the aLIGO events. 

The results of this paper are consistent with the known limits placed on the surface density of radio transients (see \citealt{murphy2013} for a summary). At the very most, CHIME or SKA1 could detect $5$ \ac{SGRB} afterglows per deg$^{2}$ per year (upper limit on the rate of \ac{BNS} coalescence) on the $\umu$Jy level, but no radio surveys have reached that sensitivity yet. Furthermore, the upper limit rate of \ac{BNS} coalescence is very unlikely. 

This work is also complementary to other work on the detectability of long GRB afterglows. \citet{ghirlanda2013} and \citet{ghirlanda2014} consider the detectability of radio afterglows from on-axis and orphan long GRBs respectively over a wide range of frequencies. Their results are more pessimistic than ours for the following reasons: the rate of long GRBs that they use is roughly a factor of 10 lower than the realistic \ac{CBC} rate; the circumburst densities that they explore are much higher than the densities we explore ($1$--$30$\,cm$^{-3}$ compared to $10^{-5}$--$1$\,cm$^{-3}$), which is appropriate for long GRBs but synchrotron self-absorption is stronger at higher densities; the microphysics parameters $\epsilon_{e}$ and $\epsilon_{B}$ that they use have lower values and would thus reduce the radio flux. This shows the sensitivity of the results on the choice of model parameters, which are motivated by and consistent with observations but remain highly uncertain. Consequently, orphan afterglow searches with radio instruments may also be able to constrain some of these parameters, such as $\theta_{\jet}$. Furthermore, long GRBs could be a background to future \ac{GW} candidate follow-ups for the pessimistic and realistic rate predictions of \ac{BNS} coalescence. Well-sampled radio lightcurves with afterglow modeling or observations at other wavelengths would be necessary to distinguish the two populations. 

\section*{Acknowledgements}
We thank Davide Burlon, Kipp Cannon, Scott Hughes, David Kaplan, Erik Katsavounidis, Tara Murphy, Alexander Urban, and Rainer Weiss for helpful discussions. We also thank Martin Bell and Andr\'{e} Offringa for answering our questions about LOFAR; Matt Dobbs, Ue-Li Pen and Keith Vanderlinde for answering our questions about CHIME; as well as Aaron Parsons and Jonathan Pober for providing us with the system parameters of HERA. This material is based upon work supported by the U.S. National Science Foundation Graduate Research Fellowship under Grant No. 1122374 and the MIT School of Science. LF and RV are supported by LIGO Laboratory. LIGO was constructed by the California Institute of Technology and Massachusetts Institute of Technology with funding from the National Science Foundation and operates under cooperative agreement PHY-0757058. Development of the Boxfit code was supported in part by NASA through grant NNX10AF62G issued through the Astrophysics Theory Program and by the NSF through grant AST-1009863. This research made use of Astropy\footnote{http://www.astropy.org}, a community-developed core Python package for Astronomy \citep{astropy2013}.

\appendix

\section{Derivation of the matched filter statistic for radio transients}
\label{appendix_match_filter}
This section derives the expression for the matched filter detection statistic used in Section~\ref{detectability} to define the detectability criteria for radio afterglows. We assume that the telescope data are already reduced to the flux time-series $x(t) = \{x_1, x_2, \cdots x_N\}$, where $x_i$ is the measured flux in the snapshot at time $t_i$  at the specific image pixel (synthesized beam) representing a particular position on the sky. The noise in these data has two components: the thermal noise, described by the Gaussian distribution with zero mean and variance $\sigma_{\thermal}$, and the classical confusion noise $c$ that is constant in time. The classical confusion noise in radio telescopes comes from unresolved point sources within the synthesized beam. While constant in time, its level varies from pixel to pixel. In the context of the search for radio transients, faint constant sources should also be included in the classical confusion noise as they can be confused with slowly varying transients or may cover them up (e.g. an unresolved radio galaxy). 

The starting point for the derivation of the detection statistic is the Neyman--Pearson optimization criteria which requires the optimal detection statistic to maximize the probability of signal detection (or efficiency) at a fixed probability of false alarm (or false positive); see \cite{neyman-1933}. The general solution of this optimization problem is the likelihood-ratio detection statistic (also known as the Bayes factor):
\begin{equation}
\label{lr_def}
\Lambda(x) = \frac{p(x \given 1)}{p(x \given 0)}\,,
\end{equation}  
where $p(x \given 1)$ is the probability of getting data $x$ when a transient signal is present, and $p(x \given 0)$ is the probability of getting the same data from noise only. For radio searches, these probability density functions are given by 
\begin{equation}\label{sig_prob_def} 
\begin{split} 
& p(x \given 1) = \\ 
& \frac{1}{\sqrt{(2\pi)^N}\sigma_{\thermal}^N} \int \mathrm{exp} \left( -\frac{1}{2}\sum_{i = 1}^{i = N} \frac{(x_i - c - Af_i)^2}{\sigma_{\thermal}^2} \right) p(c)p(A) \diff c \diff A
\end{split}
\end{equation}
\begin{equation}\label{noise_prob_def} 
p(x \given 0) = \frac{1}{\sqrt{(2\pi)^N}\sigma_{\thermal}^N} \int \mathrm{exp} \left( -\frac{1}{2}\sum_{i = 1}^{i = N} \frac{(x_i - c)^2}{\sigma_{\thermal}^2} \right) p(c) \diff c
\end{equation}
where $f(t) = A\{f_1, f_2, \cdots, f_N\}$ denotes the flux time-series of the transient signal at times $\{t_1, t_2, \cdots, t_N\}$, $A$ is the overall amplitude of the signal, $p(c)$ and $p(A)$ are the prior probability density functions for the classical confusion noise $c$ and the signal amplitude $A$ respectively. They are marginalized over, as both $c$ and $A$ are unknown. The thermal noise variance $\sigma_{\thermal}$ in each snapshot is set by the system parameters of the telescope and the snapshot integration time via \Cref{eq:thermal_noise}. For simplicity, we assume that all snapshots in the observation are of the same duration. In \Cref{sig_prob_def}, while the overall amplitude of the signal is unknown, its form is fixed by $f(t)$.

Formally, \Cref{lr_def}, \Cref{sig_prob_def}, and \Cref{noise_prob_def} define the optimal detection statistic. A standard approach is to simplify it further by evaluating the integrals in \Cref{sig_prob_def} and \Cref{noise_prob_def} using the stationary phase approximation. This is justified if $p(c)$ and $p(A)$ vary slowly relative to the exponential of Gaussian distributions. This is the case here, because both of these functions describe distributions of astrophysical sources and, thus, follow power-law distributions. 

The phase of $p(x \given 0)$,
\begin{equation}
\label{noise_phase}
S_{0} = -\frac{1}{2}\sum_{i = 1}^{i = N} \frac{(x_i - c)^2}{\sigma_{\thermal}^2} \,,
\end{equation}
has an extremum that can be computed with
\begin{equation}
\label{extrm_condition_noise}
\frac{\partial S_{0}}{\partial c} = \sum_{i = 1}^{i = N} \frac{(x_i - c)}{\sigma_{\thermal}^2} = 0 \,.
\end{equation}
The solution $c=c_0$ is given by
\begin{equation}
\label{extrm_solution_noise}
c_0 = \langle x \rangle \equiv \frac{1}{N}\sum_{i = 1}^{i = N} x_i \,.
\end{equation} 
Thus, the classical confusion noise is estimated as the average measured flux, and we approximate $p(x \given 0)$ by its value at the extremal point,
\begin{equation}
\label{noise_prob_approx}
p(x \given 0) \approx  \frac{1}{\sqrt{(2\pi)^N}\sigma_{\thermal}^N} \mathrm{exp} \left( -\frac{1}{2}\sum_{i = 1}^{i = N} \frac{(x_i - c_0)^2}{\sigma_{\thermal}^2} \right) p(c=c_0) \,.
\end{equation}

 Next, we find the extremum of the phase of $p(x \given 1)$,
\begin{equation}
\label{signal_phase}
S_{1} =  -\frac{1}{2}\sum_{i = 1}^{i = N} \frac{(x_i - c - Af_i)^2}{\sigma_{\thermal}^2}\,.
\end{equation}
The extremum conditions are
\begin{align}
\label{extrm_condition_signal_c} \frac{\partial S_{1}}{\partial c} & = \sum_{i = 1}^{i = N} \frac{(x_i - c - Af_i)}{\sigma_{\thermal}^2} = 0 \,,\\
\label{extrm_condition_signal_A} \frac{\partial S_{1}}{\partial A} & = \sum_{i = 1}^{i = N} \frac{(x_i - c - Af_i)f_i}{\sigma_{\thermal}^2} = 0 \,.
\end{align}
They are satisfied if $c = c_1$ and $A = A_{\mathrm{max}}$, given by
\begin{align}
\label{extrm_solution_signal_c}  c_1 & = \langle x \rangle - A_{\mathrm{max}} \langle f \rangle \,, \\
\label{extrm_solution_signal_A} A_{\mathrm{max}} & = \frac{(x, f - \langle f \rangle)}{(f - \langle f \rangle,  f - \langle f \rangle)} \,,
\end{align}
where the bracketed quantities define the time average, e.g. $\langle x \rangle \equiv \frac{1}{N}\sum_{i = 1}^{i = N} x_i$, and $(x,y) \equiv \sum_{i = 1}^{i = N} x_iy_i / \sigma_{\thermal}^2$ defines the inner product between any two time-series. Note that the extremal value for the classical confusion noise in the case of the signal-plus-noise hypothesis, $c_1$, is different from the extremal value of the classical confusion noise in the case of the null hypothesis, $c_0$. As previously, we approximate the probability distribution for signal-plus-noise, $p(x \given1)$, by evaluating it at its extremum,
\begin{equation}\label{signal_prob_approx}
\begin{split}
& p(x \given 1) \approx \frac{1}{\sqrt{(2\pi)^N}\sigma_{\thermal}^N} \times \\ 
& \mathrm{exp} \left( -\frac{1}{2}\sum_{i = 1}^{i = N} \frac{(x_i - c_1 - A_{\mathrm{max}}f_i)^2}{\sigma_{\thermal}^2} \right) p(c=c_1)p(A = A_{\mathrm{max}}) \,.
\end{split}
\end{equation}
Using \Cref{noise_prob_approx} and \Cref{signal_prob_approx} in the general expression for the likelihood ratio, \Cref{lr_def}, and simplifying the expressions in the exponents, we arrive at the following approximation:
\begin{equation}
\label{approx_lr}
\Lambda(x) \approx \mathrm{const} \times \mathrm{exp}\left(\frac{\rho(x)^2}{2}\right)\,,
\end{equation}
where we absorb the values of the prior probability density functions and all the terms independent of the data, $x$, into an a single, approximately constant factor, $\mathrm{const}$, and define a new quantity
\begin{equation}
\label{det_stat_def}
\rho(x) = \frac{(x, f - \langle f \rangle)}{\sqrt{(f - \langle f \rangle,  f - \langle f \rangle)}}\,.
\end{equation}
Given that the likelihood ratio, \Cref{approx_lr}, is a monotonic function of $\rho(x)$ in the stationary phase approximation, we define $\rho(x)$ to be the \textit{detection statistic} for transient radio signals.

The detection statistic $\rho(x)$ is a linear transformation of the data $x$ and can be recognized as a modified version of the well-known matched filter~\citep{helmstrom-1968}. By construction, it finds the best fit to the data by maximizing the likelihood ratio for the unknown signal amplitude $A$. The best matched value is given by \Cref{extrm_solution_signal_A} and can be re-expressed in terms of the detection statistic as
\begin{equation}
\label{best_amplitude}
A_{\mathrm{max}} = \frac{\rho(x)}{\sqrt{(f - \langle f \rangle,  f - \langle f \rangle)}}\,. 
\end{equation}
For a given measurement, $x$, the value of $\rho(x)$ reflects the likelihood for the data to contain a transient signal $A_{\mathrm{max}}f$. The detection statistic should be interpreted in the context of the probability of false alarm and  the probability of detection. After all, it is defined through maximization of the latter at a fixed value of the former. The statistical properties of $\rho(x)$ follow from its definition in \Cref{det_stat_def}. First, consider the case when the data, $x$, contain no transient signal:
\begin{equation}
\label{noise_data}
x_i = n_i + c \,,
\end{equation}
where $n_i$ is the realization of the thermal noise and $c$ is the classical confusion noise. Evaluated on these data, the detection statistic is
\begin{equation}
\label{det_stat_noise}
\rho(x) = (n, f - \langle f \rangle)\,.
\end{equation}
It is a linear superposition of Gaussian random variables $n_i$. From the properties of Gaussian random variables, the probability distribution for $\rho(x)$ in the absence of signal is also Gaussian, $p(\rho \given 0) = \mathcal{N}(\mu_0, \sigma_0)$, with mean $\mu_0 = 0$ and variance $\sigma_0 =  \sqrt{(f-\langle f \rangle,f-\langle f \rangle)}$. In an analogous manner, in the presence of the transient signal with amplitude $A^{\prime}$ in the data:
\begin{equation}
\label{signal_data}
x_i = n_i + c + A^{\prime}f_i\,,
\end{equation}
the detection statistic is given by
\begin{equation}
\label{det_stat_signal}
\rho(x) = (n, f - \langle f \rangle) + A^{\prime}(f - \langle f \rangle, f - \langle f \rangle)\,.
\end{equation}
The probability distribution for $\rho(x)$ in the presence of signal is also Gaussian, $p(\rho \given 1) = \mathcal{N}(\mu_1, \sigma_1)$, with the same variance $\sigma_1 =  \sqrt{(f-\langle f \rangle,f-\langle f \rangle)}$ but a mean shifted away from zero, $\mu_1 = A^{\prime}(f - \langle f \rangle, f - \langle f \rangle)$. From the functional form of $p(\rho \given 0)$, one can establish the threshold $\rho \ge \rho^\ast$, which would correspond to a tolerable probability of false alarm. A canonical choice for robust signal detection is to require the cumulative probability for observing the detection statistic above the threshold value $\rho^\ast$ to be less than or equal to a $5\sigma$ deviation of the random normal variable $P(\rho \ge \rho^\ast \given 0) \le [1 - \mathrm{erf}(5/\sqrt{2})] \approx 5.7 \times 10^{-7}$. Having established $\rho^\ast$ and using the expression for $p(\rho \given 1)$, one can determine the signal flux $f^\ast$ at which the signals can be detected with a high efficiency above the threshold. This in turn will define the sensitivity of the search. 

\begin{figure}
\centering
\includegraphics[width=0.5\textwidth]{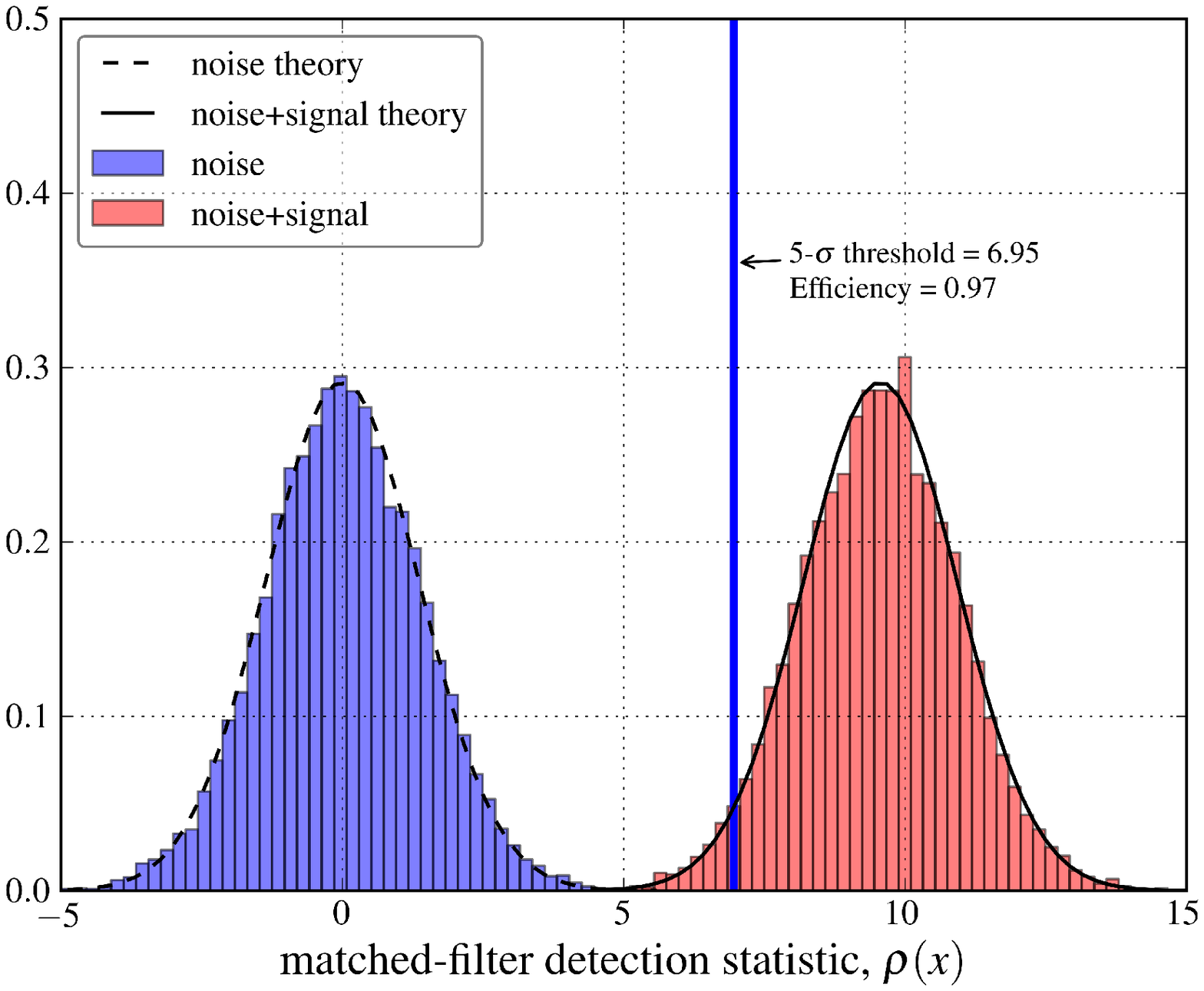}
\caption{Distributions of the detection statistic $\rho$. The plot shows the normalized histograms of $\rho$ computed for noise-only data (blue bars) and noise-plus-transient-signal data (red bars). For noise data, we generate $10^4$ realizations of the flux time-series of 1-h snapshots covering 200\,d in total duration. The thermal noise variance in each snapshot is $\sigma_{\mathrm{snapshot}} = 0.912$~mJy and the confusion noise is 10 mJy. For signal-plus-noise data, we generate another   $10^4$ noise realizations to which we add a $0.184$-mJy signal present in the first 100 days. The dashed and solid black curves are the theoretical Gaussian distributions for $p(\rho \given 0)$ and $p(\rho \given 1)$ respectively. The vertical solid line defines the $5\sigma$ false alarm probability threshold for $\rho$. In 97$\%$ of the cases, the value of $\rho$ computed from the data containing the transient signal exceeded this threshold, amounting to an efficiency of 0.97.}
\label{fig:det_stats_dist}
\end{figure}

In general, the values for $\rho^\ast$ and $f^\ast$ will depend on the form of the template $f$. It is useful, however, for our purpose to compute them using a simplified model for the template. A top-hat function is the simplest way to represent a transient:
\begin{equation}
\label{step_template}
f_i=\begin{cases}
    f_{\sref}, & \text{if $i = j$}\,,\\
    0, & \text{if $i \neq j$} \,,
  \end{cases}
\end{equation} 
where $f_{\sref}$ is the reference flux (e.g. corresponding to a source at some convenient reference distance) and $j$ labels the snapshot during which the transient is `on'. It is convenient to choose the snapshot duration to be equal to the transient duration, so that the transient is `off' during all other snapshots. Provided that the number of snapshots is $N$, the average flux for the template is $\langle f \rangle =  f_{\sref} / N$. The variance of both  $p(\rho \given 0)$ and  $p(\rho \given 1)$ is 
\begin{equation}
\label{var_simple}
\sigma_0 = \sigma_1 = \sqrt{(f-\langle f \rangle,f-\langle f \rangle)} = \frac{f_{\sref}}{\sigma_{\thermal}}\sqrt{\frac{N-1}{N}} \,,
\end{equation}
and the mean of $p(\rho \given 1)$ is 
\begin{equation}
\label{mu_signal_simple}
\mu_1 =  A^{\prime}(f - \langle f \rangle, f - \langle f \rangle) = A{\left(\frac{f_{\sref}}{\sigma_{\thermal}}\right)}^2\frac{N-1}{N}\,.
\end{equation}
To achieve a high efficiency ($\ge 97\%$) in the signal detection at $5\sigma$ probability of false alarm, we require the mean of the distribution for the signal to be at least $7\sigma$ away from the zero mean of the distribution for noise,
\begin{equation}
\label{det_condition}
\mu_1 \ge 7\sigma_0 \,,
\end{equation}
which is satisfied for the signals with the amplitude
\begin{equation}
\label{amplitude_condition}
A \ge 7\frac{\sigma_{\thermal}}{f_{\sref}}\sqrt{\frac{N}{N-1}}\,.
\end{equation}
Multiplying the amplitude by the reference flux gives us the threshold flux,
\begin{equation}
\label{thresh_flux}
f^\ast \ge A f_{\sref} = 7\sqrt{\frac{N}{N-1}}\,\sigma_{\thermal}\,.
\end{equation}
In the case of the search with two snapshots ($N=2$), the threshold on the detectable flux is $f^\ast \ge 7\sqrt{2}\sigma_{\thermal}$. In the case of infinitely many snapshots ($N \to \infty$), the threshold approaches the limit imposed by the thermal noise, $f^\ast \to 7\sigma_{\thermal}$. The effect of the constant classical confusion noise is to introduce the factor $\sqrt{N/(N-1)}$ in \Cref{thresh_flux}. In the case of long lasting transients, it is more likely that the total duration of the reference images (those without the transient) would be approximately equal to the duration of the transient, in which case the $N=2$ threshold would apply. Therefore, the price to pay for the classical confusion noise is about a 50$\%$ higher threshold on the flux as compared to the thermal limit. On the other hand, $\sigma_{\thermal}$ for long lasting transients might be quite low due to long integration times, thus making the searches for such transients quite sensitive even for the instruments limited by confusion noise. The actual threshold flux can be well below the confusion noise. \Cref{fig:det_stats_dist} illustrates this by showing the simulated distributions of the detection statistic in the presence and absence of the transient signal in the data. In the simulation, we use the noise characteristics of the MWA and generate, for each case, $10^4$ realizations of the flux time-series data of 1-h snapshots with a total duration of 200\,d. The thermal noise variance in each snapshot is $\sigma_{\mathrm{snapshot}} = 0.912$~mJy and the classical confusion noise is 10\,mJy. The flux of the transient signal in our simulations is $0.184$~mJy and the duration is 100\,d. The signal flux is computed from \Cref{thresh_flux} with $N=2$ and the variance for the thermal noise is scaled from 1\,h to 100\,d, i.e. $\sigma_{\thermal} = 0.912/\sqrt{100 \times 24} =0.0186$~mJy. As can be seen from \Cref{fig:det_stats_dist}, the efficiency in recovering such a signal with significance above $5\sigma$ is 0.97. These simulations demonstrate that, at least in principle, long faint transients can be detected well below the confusion noise. 

While simplified, this example captures the key characteristics of the search for transients and identifies its limiting factors: a sufficient number of reference images that can be used to subtract the classical confusion noise and the thermal noise. In the real-life search, a number of complications will be added to these. Imperfect calibration of the instrument may introduce another noise component that might dominate the thermal noise. Its effects can be mitigated to some degree by characterizing the calibration noise and including it in the statistical model (e.g. by modifying $\sigma_{\thermal}$ which in a more general case could correspond to a coloured Gaussian noise). As sources move on the sky, the primary beam of the instrument at the location of the source will be varying in time. Denoting the primary beam factor as $b_i$, the measured flux from the transient source as well as the classical confusion noise will be modified as $f_i \to b_if_i$ and  $c \to b_ic$ respectively. The thermal noise, on the other hand, will not be affected by the primary beam. If the primary beam is well-modelled, it can be corrected for. Omitting details, we quote the final result here. One can show that the above formalism and, in particular, the formulas for the detection statistic, \Cref{det_stat_def}, and the best-fitting signal amplitude, \Cref{best_amplitude}, still apply after the following redefinitions:
\begin{align}
\label{primary_beam_corr}
x_i &\to x_ib_i^{-1} \,,\\
\sigma_{\thermal} &\to \sigma_{\thermal} b_i^{-1}\,, \\
\langle x \rangle & \to  \sum_{i=1}^{i = N} x_iw_i \,,\\
(x,y) & \to  \sum_{i = 1}^{i = N} \frac{x_iy_i}{{(\sigma_{\thermal} b_i^{-1})}^2}\,, 
\end{align}
where $w_i = b_i^2/\sum b_j^2$, and $\langle x \rangle$ and $(x,y)$ denote the weighted average for any time-dependent quantity and the inner product between any two time-dependent quantities respectively. 

The fluctuations induced by the ionosphere and the sidelobe confusion noise may lead to transient artefacts in the data. Depending on their characteristic time-scales, these transient artefacts might give rise to a large number of false alarms and prevent one from detecting transients of the similar duration. This problem would require a special, instrument-specific treatment through developing statistical classifiers that can distinguish between artefacts and genuine signals. 

Lastly, in addition to the unknown signal amplitude, the time of arrival at the telescope is also unknown. In order to search for the transient in the data, the template $f(t)$ should be shifted in time and the detection statistic $\rho(x)$ must be re-computed for every shift. This amounts to converting the measured flux time-series, $x(t)$, into a detection statistic time-series, $\rho(x,t)$. The maximum of $\rho(x,t)$, if it crosses the detection threshold, corresponds to the most likely time of arrival of the transient. In order to search for a class of transients (e.g. of different durations or, as in the case of \ac{SGRB} afterglows, corresponding to a range of energies, circumburst densities, and observer angles), a bank of templates with various lightcurves is required. Broadening the search in the space of transient signals allows one to detect different types of transients but at the same time increases the probability of false alarm. Searching for transients in time and in image pixels and extending the bank of templates increase the number of independent statistical trials that must be accounted for when calculating the probability of false alarm and defining the detection threshold. In practice, the trials factor is typically estimated from simulations or using a subset of the data identified as the `playground'. Here we just remark that, for example, in order to account for $10^7$ independent trials in a search, it is sufficient to increment the detection threshold in \Cref{thresh_flux} by $2.5 \sigma_{\thermal}$.

\section{Thermal Sensitivity of CHIME and HERA}\label{appendix_chime_sens}

CHIME is a drift-scan telescope with cylindrical reflectors oriented in the north-south direction without any moving parts. Its instantaneous field of view can be approximated as a 2.5$\degr$ narrow band spanning the whole sky from north to south. As the Earth rotates, the telescope observes effectively half of the sky every day. The integration time of a source is a function of its declination. As a result, the instrument sensitivity will vary with source declination. In order to account for this, we compute the integration time as a function of declination $t_{\sint}(\delta)$.

\begin{figure}
  \centering
  \includegraphics[width=0.5\textwidth]{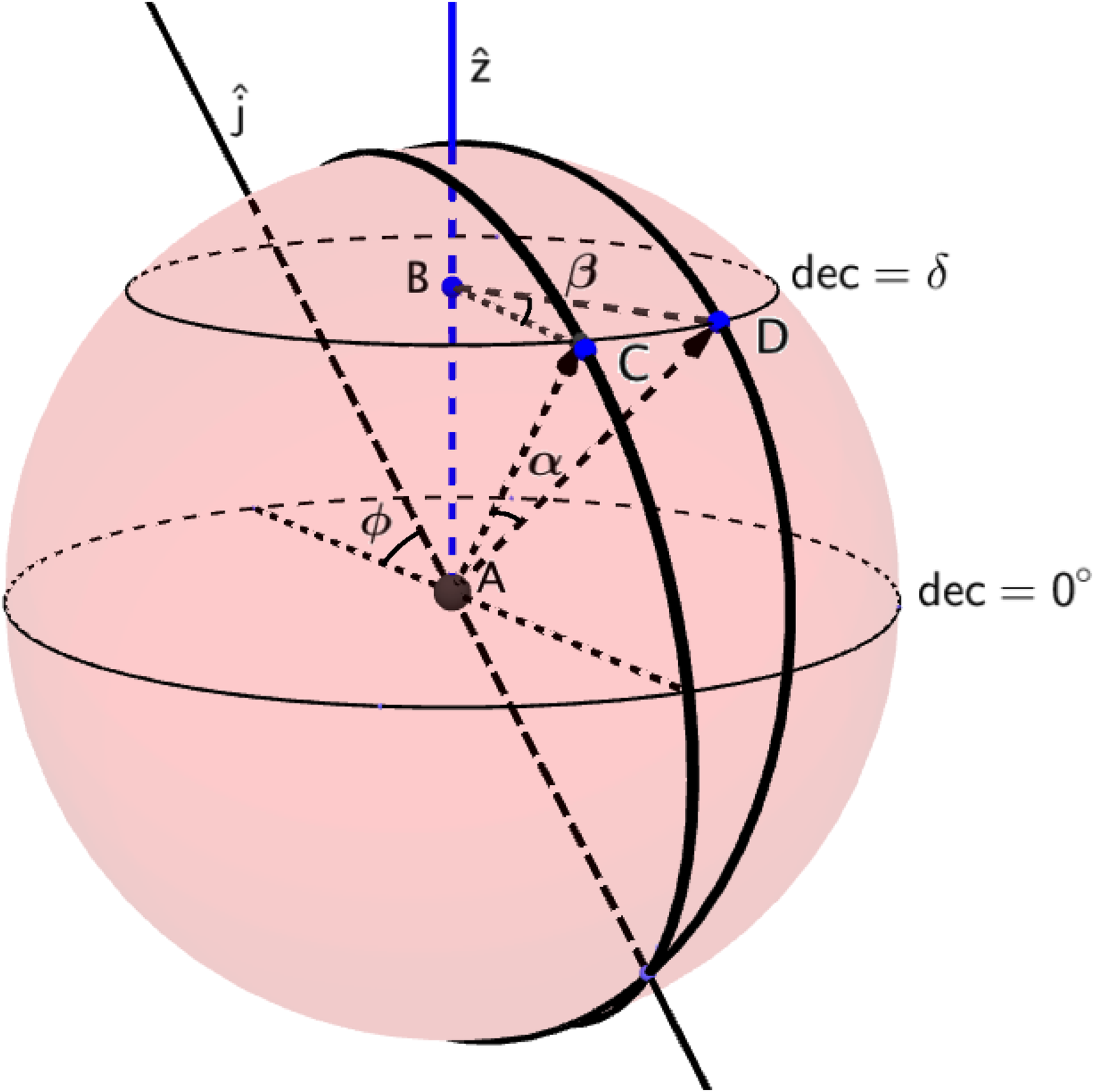}
  \caption{Visualization of CHIME observations. The CHIME field of view is defined by two planes intersecting at an angle $\Delta=2.5 \degr$, where the line of intersection, shown as axis $\hat{j}$, is oriented in the north-south direction and makes an angle $\phi = 45\degr$ with the $z$-axis. The intersections of these planes with the celestial sphere, indicated by the thick solid arcs, defines the instantaneous field of view. As the Earth rotates, sources come in and out of the field of view, and the typical daily trajectory of a source at $\mathrm{Dec}=\delta$ is shown as the circle through points $C$ and $D$. This source is observable only when it is located on the short arc between points $C$ and $D$, and the ratio of the angle subtended by this arc, $\beta = \protect\angle CBD$, to 2$\pi$ gives the fraction of the day during which this source is observable. The total exposure time thus increases with $\delta$ in the northern hemisphere; the sources in the southern hemisphere are either observable for only a very short time or inaccessible. The sources with $\delta > 45\degr$ are observed twice a day, and the sources with $\delta \ga 88.75\degr$ are always in the field of view.}
  \label{fig:chime_fov}
\end{figure}

The instantaneous field of view of CHIME is modeled by two intersecting planes with the angle between their normal directions $\Delta \approx 2.5 \degr$ defining the aperture in the east-west direction. The slice of the celestial sphere between the two planes defines the instantaneous field of view of the telescope. \Cref{fig:chime_fov} shows the visualization of the field of view of CHIME, which we assume to be located at a latitude $\phi=45\degr$.  

The integration time during a single pass of the source across the field of view is proportional to the angle $\beta$ subtended by the arc between the planes, designated as $\beta = \angle CBD$ on \Cref{fig:chime_fov}. When expressed as a fraction of the full 24-h day (corresponding to a single complete revolution), the integration time is given by
\begin{equation}\label{initial_t_exp}
  t_{\sint} = \frac{\beta}{2\pi}\,\text{day}\,.
\end{equation}
Using the fact that triangles $CBD$ and $CAD$ share the same side, CD, one can express $\beta = \angle CBD$ in terms of the angle $\alpha = \angle CAD$ and the source declination $\delta$:
\begin{equation}\label{cosbeta}
  \cos\beta = 1 - \frac{1 - \cos\alpha}{\cos^2\delta}\,.
\end{equation}
In deriving this expression, we use $BD=BC$, $AC=AD$ and $BC = AC\cos\delta$. Next, we find $\cos\alpha$ by computing the scalar product between the two unit vectors $\bmath{a}_{1} = \bmath{AC}/|\bmath{AC}|$ and $\bmath{a}_2=\bmath{AD}/|\bmath{AD}|$. These vectors lie in the planes defining the field of view. The angle between the planes is $\Delta$. Vectors $\bmath{a}_{1}$ and $\bmath{a}_{2}$ are identical up to a rotation by $\Delta$ around the axis defined as the line of intersection of the planes, axis $\hat{j}$ in \Cref{fig:chime_fov}. Taking advantage of this fact, we find:
\begin{equation}\label{cosalpha}
  \cos \alpha = (\bmath{a}_{1},\bmath{a}_{2}) = \sin^2(\phi - \delta) + \cos\Delta \cos^2(\phi - \delta)\,.
\end{equation}
Using \Cref{cosalpha} and \Cref{cosbeta} in \Cref{initial_t_exp}, we arrive at the final expression for the integration time in a single transit:
\begin{equation}\label{final_t_exp}
  t_{\sint}(\delta) = \frac{1}{2\pi} \arccos\left[ \frac{-\sin^2\delta + \sin^2(\phi - \delta) + \cos\Delta \cos^2(\phi - \delta) }{\cos^2\delta}\right]\,.
\end{equation}
The problem of calculating the integration time for a cylindrical telescope with an arbitrary location and orientation was treated in \cite{Moniez:2012iy}. Our derivation is different from the approach used in \cite{Moniez:2012iy}. As a result, our formula for $t_{\sint}$ given by \Cref{final_t_exp} does not match in its functional form the formula given in \cite{Moniez:2012iy}. However, we verified numerically that both formulas, ours and \cite{Moniez:2012iy}, lead to identical results. 

The sources south of the telescope, $\delta \le \phi$, are observed once a day. The sources with declinations greater than the latitude of the telescope, $\delta > \phi$, transit twice a day through the field of view of the telescope. The second pass occurs through a different part of the field of view (located on the opposite side of the northern hemisphere as shown on \Cref{fig:chime_fov}) and, therefore, its integration time is not the same as the one for the first transit. The integration time for the second pass can be computed by replacing $\phi - \delta$ with $\phi +\delta$ in \Cref{final_t_exp}. The total integration time for such a source is the sum of the two integration times. Finally, the sources that are very near the north pole ($\delta \ga 88.75\degr$) are always in the field of view of the telescope. 

Substituting \Cref{final_t_exp} into \Cref{eq:thermal_noise}, we can now compute the thermal noise as a function of $\delta$:
\begin{equation}\label{thermal_noise_formula}
  \sigma_{\mathrm{CHIME}}(\delta) = \left( \frac{2k_{\mathrm{B}} T_{\mathrm{sys}}}{A_{\mathrm{eff}} N_{\mathrm{ant}} \epsilon_{c}} \right) \frac{1}{\sqrt{N_{\mathrm{pol}}B \,t_{\sint}(\delta)}}\,,
\end{equation}
where we set $T_{\mathrm{sys}}=100$\,K, $\epsilon_{c}=1.0$, $N_{\mathrm{pol}} = 2.0$, $B = 400$\,MHz, and $A_{\mathrm{total}} = A_{\mathrm{eff}} N_{\mathrm{ant}} = 1500\,$m$^2$ for the CHIME pathfinder and $A_{\mathrm{total}} = A_{\mathrm{eff}} N_{\mathrm{ant}} = 10000\,$m$^2$ for the full CHIME. The latitude is set to $\phi = 45 \degr$ and the opening angle $\Delta = 2.5 \degr$. When averaged over all declinations in the northern hemisphere $0\degr \leq \delta \leq 90\degr,$ the average thermal noise in a 1-day observation is
\begin{align}
  \left\langle \sigma_{\mathrm{CHIME\,path}} \right\rangle &=  0.240 \text{ mJy\,, and}  \label{noise_chime_path} \\
  \left\langle\sigma_{\mathrm{CHIME}} \right\rangle   &=  0.036 \text{ mJy}\,, \label{noise_chime_full} 
\end{align}  
for the CHIME pathfinder and the full CHIME, respectively.

When estimating the expected number of detections for CHIME, we need to average the horizon distance in \Cref{eq:detection_rate} over all declinations. Since $d_{\mathrm{H}} \propto \sigma_{\thermal}^{-1/2}$, one should average the noise taken in the power of $-3/2$, $\left\langle \sigma_{\mathrm{CHIME}}^{(-3/2)} \right\rangle$, which differs from the average noise taken in the same power by a factor of 1.19,
\begin{equation}\label{chime_corr_factor}
  \left\langle \sigma_{\mathrm{CHIME}}^{(-3/2)} \right\rangle = 1.19 {\left\langle \sigma_{\mathrm{CHIME}} \right\rangle}^{(-3/2)}\,.
\end{equation}
In our calculation of the expected number of detections for CHIME, we use \Cref{noise_chime_path} for the CHIME pathfinder and \Cref{noise_chime_full} for the full CHIME in \Cref{eq:detection_rate}, and then correct the results by multiply them by the factor of 1.19. 

HERA is a drift-scan telescope with a $8.7\degr$~primary beam full-width at half-maximum. As it will be located in South Africa, we assume the telescope to be at the latitude $\phi=-30\degr$. Setting $\Delta=8.7\degr$, we estimate $t_{\sint}$ for a source at zenith ($\delta=\phi$) using \Cref{final_t_exp}. We neglect the effects related to the circular shape of the field of view, which should be small. Because the area surveyed by the telescope is relatively narrow in the north-south direction, we can neglect the dependence of $t_{\sint}$ on $\delta$. The resulting $t_{\sint}$ for the sources observed by HERA in a single drift-scan observation is $t_{\sint} = 0.67$\,h. We compute the survey area accessible to the telescope by approximating it as a strip of sky centred on $\delta = -30\degr$ with width $8.7\degr$: $\Omega_{\mathrm{HERA}} = 2712.4$\,deg$^{2}$.  Lastly, when computing the thermal noise for HERA, we set $T_{\mathrm{sys}}=351$\,K, $\epsilon_{c}=1.0$, $N_{\mathrm{pol}} = 2.0$, $B = 100$\,MHz, and $A_{\mathrm{total}} = N_{\mathrm{ant}} \pi r_{\mathrm{ant}}^2 =  84204$~m$^2$, where the antenna radius is $r_{\mathrm{ant}}= 7$~m. Using all of this in \Cref{thermal_noise_formula}, we estimate the thermal noise for HERA in a 1-day drift scan observation to be
\begin{align}
  \sigma_{\mathrm{HERA}} &=  0.017 \text{ mJy}  \label{hera}\,.
\end{align}

\section{Estimation of the fraction of gravitational wave events detectable in radio follow-up observations}\label{appendix_ligo_followup}

The necessary condition for both the \ac{GW} and afterglow radio emission to be detectable is for the \ac{CBC} to be within the detectability range of both instruments. The reach of the \ac{GW} detectors can be expressed in terms of the horizon distance, $D^{\mathrm{GW}}_{\mathrm{horizon}}$ -- the distance at which the signal from an optimally-located and oriented \ac{CBC} produces an event with the matched filter \ac{SNR} of 8 in the detector. In the nominal regime, this condition corresponds to a $5\sigma$ detection. The \ac{SNR} of a \ac{GW} signal from the \ac{CBC} is proportional to its amplitude, which in turn is inversely proportional to the physical distance to the \ac{CBC} and depends on the location and the orientation of the binary relative to the detector. All of these parameters can be combined into a single factor called the effective distance, $D_{\mathrm{eff}}$:
\begin{equation}
\label{D_eff_def}
D_{\mathrm{eff}} = d{\left[F_{+}^2(\theta, \phi, \psi){\left(\frac{1+\cos^2\iota}{2} \right)}^2 + F_{\times}^2(\theta, \phi, \psi) \cos^2\iota \right]}^{-1/2}\,,
\end{equation}
where $d$ is the physical distance to the \ac{CBC}, the polar angles $(\theta, \phi)$ define the position of the source on the sky in the detector coordinate system (for a detector with orthogonal arms, the $x$- and $y$-axes are aligned with the arms of the detector and the $z$-axis is perpendicular to the plane of the detector), $0 \leq \psi \leq 2\pi$ is the angle describing the polarization of the \ac{GW} event, the inclination angle $\iota$ is the angle between the normal vector to the binary disc and the vector pointing to the detector (the disc is face-on when $\iota = 0 \degr$ and edge-on when $\iota = 90 \degr$), and $F_{+}(\theta, \phi, \psi)$ and $F_{\times}(\theta, \phi, \psi)$ are the \ac{GW} detector antenna beam patterns given by 
\begin{align}
F_{+}(\theta, \phi, \psi) & = -\frac{1}{2}(1 + \cos^2 \theta)\cos 2\phi \cos 2\psi - \cos \theta \sin 2\phi \sin 2\psi\,, \\
F_{\times}(\theta, \phi, \psi) & = +\frac{1}{2}(1 + \cos^2 \theta)\cos 2\phi \sin 2\psi - \cos \theta \sin 2\phi \cos 2\psi\,.
\end{align}
For an optimally-located and oriented binary, $\iota = 0 \degr$ and $F_{+}^2 + F_{\times}^2 = 1$, in which case $D_{\mathrm{eff}} = d$. In all other cases, $D_{\mathrm{eff}}  > d$. Given that $\mathrm{\ac{SNR}} \propto 1/D_{\mathrm{eff}}$, the effective distance determines the strength of the \ac{GW} signal from a \ac{CBC} with an arbitrary location and orientation relative to the optimally-located and oriented \ac{CBC}. Applying the condition used to define $D^{\mathrm{GW}}_{\mathrm{horizon}}$ to a generic \ac{CBC},
\begin{equation}
D_{\mathrm{eff}} \le D^{\mathrm{GW}}_{\mathrm{horizon}} \,,
\end{equation}
 and solving for $d$, we find the reach of a \ac{GW} detector as a function of the \ac{CBC} location and orientation:
 \begin{equation} \label{gw_reach}
\begin{split}
 d_{\mathrm{GW}}(\theta, \phi, \psi, \iota) = & D^{\mathrm{GW}}_{\mathrm{horizon}} \times \\
 & {\left[F_{+}^2(\theta, \phi, \psi){\left(\frac{1+\cos^2\iota}{2} \right)}^2 + F_{\times}^2(\theta, \phi, \psi) \cos^2\iota \right]}^{1/2}\,.
\end{split}
 \end{equation}
From the definition of the effective distance,  $d_{\mathrm{GW}}$ is always less than or equal to  $D^{\mathrm{GW}}_{\mathrm{horizon}}$. \footnote{Another often quoted measure of sensitivity of \ac{GW} detectors is the inspiral range distance, $d_{\mathrm{range}}$, defined by $d_{\mathrm{range}}^3 = \langle d_{\mathrm{GW}}^3 \rangle$, where the averaging is done over all possible locations and orientations. The inspiral range distance is related to the horizon distance by $d_{\mathrm{range}} = D^{\mathrm{GW}}_{\mathrm{horizon}}/2.26$. The inspiral range is used to compute the average sensitive volume of the \ac{GW} detector and the expected rate of detections.}

For a given circumburst density, the reach of a radio telescope is characterized by $d_{\mathrm{H}}(\theta_{\obs}, E_{\iso})$, which depends on the orientation of the jet and the energy. For a \ac{CBC} to be detected by a \ac{GW} detector and a radio telescope, it must be within the reach of both instruments. $E_{\iso}$ should not be correlated with any of the extrinsic parameters of the binary $(\theta, \phi, \psi, \iota)$. Depending on the formation mechanism, the jet orientation $\theta_{\obs}$ may or may not be correlated with the inclination of the binary, $\iota$. In order to encompass all possibilities, we consider two limiting cases: (i) $\theta_{\obs} = \iota$, (ii) $\theta_{\obs}$ and $\iota$ are completely uncorrelated. Averaging over all parameters describing the \ac{CBC} and the SGRB jet, we find that the average fraction of the \ac{GW} events that can also be detected by a radio telescope in case (i) is
\begin{equation}\label{radio_to_gw_frac_corr}
\begin{split}
\left \langle \frac{N_{\mathrm{radio}}}{N_{\mathrm{GW}}} \right \rangle_{(\mathrm{i})} = & \left \langle \frac{d_{\mathrm{H}}^3(\iota, E_{\iso})}{d_{\mathrm{GW}}^3(\theta, \phi, \psi, \iota)}\Theta\left(1 - \frac{d_{\mathrm{H}}(\iota, E_{\iso})}{d_{\mathrm{GW}}(\theta, \phi, \psi, \iota)}\right)\right. + \\
& \left.\Theta\left( \frac{d_{\mathrm{H}}(\iota, E_{\iso})}{d_{\mathrm{GW}}(\theta, \phi, \psi, \iota)} - 1\right)\right \rangle_{(E_{\iso}, \theta, \phi, \psi, \iota)}\,,
\end{split}
\end{equation}
and in case (ii) is
\begin{equation}\label{radio_to_gw_frac_uncorr}
\begin{split}
\left \langle \frac{N_{\mathrm{radio}}}{N_{\mathrm{GW}}} \right \rangle_{(\mathrm{ii})} = & \left \langle \frac{d_{\mathrm{H}}^3(\theta_{\obs}, E_{\iso})}{d_{\mathrm{GW}}^3(\theta, \phi, \psi, \iota)}\Theta\left(1 - \frac{d_{\mathrm{H}}(\theta_{\obs}, E_{\iso})}{d_{\mathrm{GW}}(\theta, \phi, \psi, \iota)}\right)\right. + \\
& \left. \Theta\left( \frac{d_{\mathrm{H}}(\theta_{\obs}, E_{\iso})}{d_{\mathrm{GW}}(\theta, \phi, \psi, \iota)} - 1\right)\right \rangle_{(\theta_{\obs},E_{\iso}, \theta, \phi, \psi, \iota)}\,.
\end{split}
\end{equation}
Note that in case (i), \Cref{radio_to_gw_frac_corr}, we explicitly impose the condition $\theta_{\obs} = \iota$ and perform the averaging over a reduced set of parameters. The Heaviside step functions in \Cref{radio_to_gw_frac_corr} and \Cref{radio_to_gw_frac_uncorr} impose the condition for the ratio of the reach of the radio telescope and the \ac{GW} detector not to exceed one. The \ac{CBC} with $d_{\mathrm{GW}} < d <  d_{\mathrm{H}} $ can be detected only as a radio afterglow. Thus, in \Cref{radio_to_gw_frac_corr} and \Cref{radio_to_gw_frac_uncorr}, we set $d_{\mathrm{H}}/d_{\mathrm{GW}}=1$ whenever $d_{\mathrm{H}}$ exceeds $d_{\mathrm{GW}}$.   

Setting $D^{\mathrm{GW}}_{\mathrm{horizon}} = 450$\,Mpc (corresponding to designed aLIGO sensitivity for \ac{BNS}) in \Cref{gw_reach}, and substituting the result in \Cref{radio_to_gw_frac_corr} and \Cref{radio_to_gw_frac_uncorr}, we compute the expected fraction of aLIGO events to be detectable by various radio telescopes.  

For each telescope that we consider, we allow a full year of observations that is split evenly between aLIGO events. The total number of independent observations that will be required to follow up all accessible \ac{GW} events is determined by two factors: the rate of \ac{GW} events and the number of pointings necessary to cover the region of sky localization uncertainty of \ac{GW} detectors.  When computing the total number of observations, we account for the fact that the radio telescopes can typically access only half of the sky, which reduces the number of the \ac{GW} events that can be followed up by half. We take a typical uncertainty in the localization of a \ac{CBC} signal with the aLIGO-aVirgo network to be $100$ deg$^2$. If a telescope required multiple pointings to cover such a region, we increase the number of observations for that telescope accordingly. Most of the low frequency radio telescopes have a sufficiently large field of view to cover the localization region of a \ac{GW} source with a single pointing. Knowing the approximate locations of \ac{GW} sources allows one to reduce the total area of the sky that needs to be observed. This gives advantage to the targeted follow-up observations over a blind, all-sky survey. However, if the density of \ac{GW} events is high, a widefield telescope might end up covering the entire accessible sky in the process of the follow-up observations. It would  be equivalent to performing the blind, all-sky survey. Thus, we set the total number of observations to be either the number of pointings required to follow up all \ac{GW} events or the total of number of pointings necessary to cover 2$\pi$ of the sky, whichever is smaller. The total observation time (1\,yr) is divided evenly between the follow-up observations. In calculating $\sigma_{\thermal}$, we set $t_{\sint}$ to be half of the time allocated for each follow-up observation. As before, we allocate half of the single observation time for reference imaging, which is required to detect transients. We calculate $d_{\mathrm{H}}$ following the same procedure as in the case of a blind survey in Section~\ref{blind_survey}. Evaluating the multi-dimensional integrals in  \Cref{radio_to_gw_frac_corr} and \Cref{radio_to_gw_frac_uncorr} numerically, we compute the average fraction of aLIGO events detectable in the radio follow-up observations for each telescope and three different circumburst densities, $n=10^{-5}, 10^{-3}, 1.0$\, cm$^{-3}$. The results are listed in \Cref{tb:ligo_followup}.

\acrodef{BNS}{binary neutron stars}
\acrodef{CBC}{compact binary coalescence}
\acrodef{EM}{electromagnetic}
\acrodef{GW}{gravitational wave}
\acrodef{SGRB}{short gamma-ray bursts}
\acrodef{SNR}{signal-to-noise ratio}

\bibliographystyle{mn2e}
\bibliography{boxfit}

\begin{thebibliography}{}

\bibitem[\protect\citeauthoryear{{Abadie}, {Abbott}, {Abbott}, {Abernathy},
  {Accadia}, {Acernese}, {Adams}, {Adhikari}, {Ajith}, {Allen} \& et
  al.}{{Abadie} et~al.}{2010}]{abadie2010}
{Abadie} J.,  {Abbott} B.~P.,  {Abbott} R.,  {Abernathy} M.,  {Accadia} T.,
  {Acernese} F.,  {Adams} C.,  {Adhikari} R.,  {Ajith} P.,  {Allen} B.,    et
  al. 2010, Classical and Quantum Gravity, 27, 173001

\bibitem[\protect\citeauthoryear{{Allen}, {Anderson}, {Brady}, {Brown} \&
  {Creighton}}{{Allen} et~al.}{2012}]{Allen:2005fk}
{Allen} B.,  {Anderson} W.~G.,  {Brady} P.~R.,  {Brown} D.~A.,    {Creighton}
  J.~D.~E.,  2012, Phys. Rev. D, 85, 122006

\bibitem[\protect\citeauthoryear{{Anderson}, {van der Horst}, {Staley},
  {Fender}, {Wijers}, {Scaife}, {Rumsey}, {Titterington}, {Rowlinson} \&
  {Saunders}}{{Anderson} et~al.}{2014}]{anderson2014}
{Anderson} G.~E.,  {van der Horst} A.~J.,  {Staley} T.~D.,  {Fender} R.~P.,
  {Wijers} R.~A.~M.~J.,  {Scaife} A.~M.~M.,  {Rumsey} C.,  {Titterington}
  D.~J.,  {Rowlinson} A.,    {Saunders} R.~D.~E.,  2014, arXiv: 1403.2217

\bibitem[\protect\citeauthoryear{{Arun}, {Tagoshi}, {Kant Mishra} \&
  {Pai}}{{Arun} et~al.}{2014}]{arun2014}
{Arun} K.~G.,  {Tagoshi} H.,  {Kant Mishra} C.,    {Pai} A.,  2014, arXiv:
  1403.6917

\bibitem[\protect\citeauthoryear{{Astropy Collaboration}, {Robitaille},
  {Tollerud}, {Greenfield}, {Droettboom}, {Bray}, {Aldcroft}, {Davis},
  {Ginsburg}, {Price-Whelan} \& et al.}{{Astropy Collaboration}
  et~al.}{2013}]{astropy2013}
{Astropy Collaboration} {Robitaille} T.~P.,  {Tollerud} E.~J.,  {Greenfield}
  P.,  {Droettboom} M.,  {Bray} E.,  {Aldcroft} T.,  {Davis} M.,  {Ginsburg}
  A.,  {Price-Whelan} A.~M.,    et al. 2013, A\&A, 558, A33

\bibitem[\protect\citeauthoryear{{Bell}, {Murphy}, {Kaplan}, {Hancock},
  {Gaensler}, {Banyer}, {Bannister}, {Trott}, {Hurley-Walker}, {Wayth} \& et
  al.}{{Bell} et~al.}{2014}]{bell2014}
{Bell} M.~E.,  {Murphy} T.,  {Kaplan} D.~L.,  {Hancock} P.,  {Gaensler} B.~M.,
  {Banyer} J.,  {Bannister} K.,  {Trott} C.,  {Hurley-Walker} N.,  {Wayth}
  R.~B.,    et al. 2014, MNRAS, 438, 352

\bibitem[\protect\citeauthoryear{{Berger}}{{Berger}}{2009}]{berger2009}
{Berger} E.,  2009, ApJ, 690, 231

\bibitem[\protect\citeauthoryear{{Berger}}{{Berger}}{2013}]{berger2013_review}
{Berger} E.,  2013, arXiv: 1311.2603

\bibitem[\protect\citeauthoryear{{Berger}, {Fong} \& {Chornock}}{{Berger}
  et~al.}{2013}]{berger2013}
{Berger} E.,  {Fong} W.,    {Chornock} R.,  2013, ApJ, 774, L23

\bibitem[\protect\citeauthoryear{{Berger}, {Price}, {Cenko}, {Gal-Yam},
  {Soderberg}, {Kasliwal}, {Leonard}, {Cameron}, {Frail}, {Kulkarni} \& et
  al.}{{Berger} et~al.}{2005}]{berger2005}
{Berger} E.,  {Price} P.~A.,  {Cenko} S.~B.,  {Gal-Yam} A.,  {Soderberg} A.~M.,
   {Kasliwal} M.,  {Leonard} D.~C.,  {Cameron} P.~B.,  {Frail} D.~A.,
  {Kulkarni} S.~R.,    et al. 2005, Nature, 438, 988

\bibitem[\protect\citeauthoryear{Biswas, Brady, Burguet-Castell, Cannon,
  Clayton, Dietz, Fotopoulos, Goggin, Keppel, Pankow, Price \& Vaulin}{Biswas
  et~al.}{2012}]{PhysRevD.85.122008}
Biswas R.,  Brady P.~R.,  Burguet-Castell J.,  Cannon K.,  Clayton J.,  Dietz
  A.,  Fotopoulos N.,  Goggin L.~M.,  Keppel D.,  Pankow C.,  Price L.~R.,
  Vaulin R.,  2012, Phys. Rev. D, 85, 122008

\bibitem[\protect\citeauthoryear{{Blandford} \& {McKee}}{{Blandford} \&
  {McKee}}{1976}]{blandford1976}
{Blandford} R.~D.,  {McKee} C.~F.,  1976, Physics of Fluids, 19, 1130

\bibitem[\protect\citeauthoryear{{Bloom}, {Holz}, {Hughes}, {Menou}, {Adams},
  {Anderson}, {Becker}, {Bower}, {Brandt}, {Cobb} \& et al.}{{Bloom}
  et~al.}{2009}]{bloom2009}
{Bloom} J.~S.,  {Holz} D.~E.,  {Hughes} S.~A.,  {Menou} K.,  {Adams} A.,
  {Anderson} S.~F.,  {Becker} A.,  {Bower} G.~C.,  {Brandt} N.,  {Cobb} B.,
  et al. 2009, arXiv: 0902.1527

\bibitem[\protect\citeauthoryear{{Burrows}, {Grupe}, {Capalbi}, {Panaitescu},
  {Patel}, {Kouveliotou}, {Zhang}, {M{\'e}sz{\'a}ros}, {Chincarini}, {Gehrels}
  \& {Wijers}}{{Burrows} et~al.}{2006}]{burrows2006}
{Burrows} D.~N.,  {Grupe} D.,  {Capalbi} M.,  {Panaitescu} A.,  {Patel} S.~K.,
  {Kouveliotou} C.,  {Zhang} B.,  {M{\'e}sz{\'a}ros} P.,  {Chincarini} G.,
  {Gehrels} N.,    {Wijers} R.~A.~M.,  2006, ApJ, 653, 468

\bibitem[\protect\citeauthoryear{{Chandra} \& {Frail}}{{Chandra} \&
  {Frail}}{2012}]{chandra2012}
{Chandra} P.,  {Frail} D.~A.,  2012, ApJ, 746, 156

\bibitem[\protect\citeauthoryear{{Condon}}{{Condon}}{1974}]{condon1974}
{Condon} J.~J.,  1974, ApJ, 188, 279

\bibitem[\protect\citeauthoryear{{Cordes}, {Lazio} \& {McLaughlin}}{{Cordes}
  et~al.}{2004}]{cordes2004}
{Cordes} J.~M.,  {Lazio} T.~J.~W.,    {McLaughlin} M.~A.,  2004, New Astronomy
  Reviews, 48, 1459

\bibitem[\protect\citeauthoryear{{Dalal}, {Holz}, {Hughes} \& {Jain}}{{Dalal}
  et~al.}{2006}]{dalal2006}
{Dalal} N.,  {Holz} D.~E.,  {Hughes} S.~A.,    {Jain} B.,  2006, Phys. Rev. D,
  74, 063006

\bibitem[\protect\citeauthoryear{{Eichler}, {Livio}, {Piran} \&
  {Schramm}}{{Eichler} et~al.}{1989}]{eichler1989}
{Eichler} D.,  {Livio} M.,  {Piran} T.,    {Schramm} D.~N.,  1989, Nature, 340,
  126

\bibitem[\protect\citeauthoryear{{Ellingson}, {Taylor}, {Craig}, {Hartman},
  {Dowell}, {Wolfe}, {Clarke}, {Hicks}, {Kassim}, {Ray}, {Rickard}, {Schinzel}
  \& {Weiler}}{{Ellingson} et~al.}{2013}]{ellingson2013}
{Ellingson} S.~W.,  {Taylor} G.~B.,  {Craig} J.,  {Hartman} J.,  {Dowell} J.,
  {Wolfe} C.~N.,  {Clarke} T.~E.,  {Hicks} B.~C.,  {Kassim} N.~E.,  {Ray}
  P.~S.,  {Rickard} L.~J.,  {Schinzel} F.~K.,    {Weiler} K.~W.,  2013, IEEE
  Transactions on Antennas and Propagation, 61, 2540

\bibitem[\protect\citeauthoryear{{Fong}, {Berger}, {Chornock}, {Tanvir},
  {Levan}, {Fruchter}, {Graham}, {Cucchiara} \& {Fox}}{{Fong}
  et~al.}{2011}]{fong2011}
{Fong} W.,  {Berger} E.,  {Chornock} R.,  {Tanvir} N.~R.,  {Levan} A.~J.,
  {Fruchter} A.~S.,  {Graham} J.~F.,  {Cucchiara} A.,    {Fox} D.~B.,  2011,
  ApJ, 730, 26

\bibitem[\protect\citeauthoryear{{Fong}, {Berger} \& {Fox}}{{Fong}
  et~al.}{2010}]{fong2010}
{Fong} W.,  {Berger} E.,    {Fox} D.~B.,  2010, ApJ, 708, 9

\bibitem[\protect\citeauthoryear{{Fong}, {Berger}, {Margutti}, {Zauderer},
  {Troja}, {Czekala}, {Chornock}, {Gehrels}, {Sakamoto}, {Fox} \&
  {Podsiadlowski}}{{Fong} et~al.}{2012}]{fong2012}
{Fong} W.,  {Berger} E.,  {Margutti} R.,  {Zauderer} B.~A.,  {Troja} E.,
  {Czekala} I.,  {Chornock} R.,  {Gehrels} N.,  {Sakamoto} T.,  {Fox} D.~B.,
  {Podsiadlowski} P.,  2012, ApJ, 756, 189

\bibitem[\protect\citeauthoryear{{Fong}, {Berger}, {Metzger}, {Margutti},
  {Chornock}, {Migliori}, {Foley}, {Zauderer}, {Lunnan}, {Laskar}, {Desch},
  {Meech}, {Sonnett}, {Dickey}, {Hedlund} \& {Harding}}{{Fong}
  et~al.}{2014}]{fong2014}
{Fong} W.,  {Berger} E.,  {Metzger} B.~D.,  {Margutti} R.,  {Chornock} R.,
  {Migliori} G.,  {Foley} R.~J.,  {Zauderer} B.~A.,  {Lunnan} R.,  {Laskar} T.,
   {Desch} S.~J.,  {Meech} K.~J.,  {Sonnett} S.,  {Dickey} C.,  {Hedlund} A.,
   {Harding} P.,  2014, ApJ, 780, 118

\bibitem[\protect\citeauthoryear{{Frail}, {Kulkarni}, {Ofek}, {Bower} \&
  {Nakar}}{{Frail} et~al.}{2012}]{frail2012}
{Frail} D.~A.,  {Kulkarni} S.~R.,  {Ofek} E.~O.,  {Bower} G.~C.,    {Nakar} E.,
   2012, ApJ, 747, 70

\bibitem[\protect\citeauthoryear{{Gal-Yam}, {Ofek}, {Poznanski}, {Levinson},
  {Waxman}, {Frail}, {Soderberg}, {Nakar}, {Li} \& {Filippenko}}{{Gal-Yam}
  et~al.}{2006}]{galyam2006}
{Gal-Yam} A.,  {Ofek} E.~O.,  {Poznanski} D.,  {Levinson} A.,  {Waxman} E.,
  {Frail} D.~A.,  {Soderberg} A.~M.,  {Nakar} E.,  {Li} W.,    {Filippenko}
  A.~V.,  2006, ApJ, 639, 331

\bibitem[\protect\citeauthoryear{{Ghirlanda}, {Burlon}, {Ghisellini},
  {Salvaterra}, {Bernardini}, {Campana}, {Covino}, {D'Avanzo}, {D'Elia},
  {Melandri}, {Murphy}, {Nava}, {Vergani} \& {Tagliaferri}}{{Ghirlanda}
  et~al.}{2014}]{ghirlanda2014}
{Ghirlanda} G.,  {Burlon} D.,  {Ghisellini} G.,  {Salvaterra} R.,  {Bernardini}
  M.~G.,  {Campana} S.,  {Covino} S.,  {D'Avanzo} P.,  {D'Elia} V.,  {Melandri}
  A.,  {Murphy} T.,  {Nava} L.,  {Vergani} S.~D.,    {Tagliaferri} G.,  2014,
  arXiv: 1402.6338

\bibitem[\protect\citeauthoryear{{Ghirlanda}, {Salvaterra}, {Burlon},
  {Campana}, {Melandri}, {Bernardini}, {Covino}, {D'Avanzo}, {D'Elia},
  {Ghisellini}, {Nava}, {Prandoni}, {Sironi}, {Tagliaferri}, {Vergani} \&
  {Wolter}}{{Ghirlanda} et~al.}{2013}]{ghirlanda2013}
{Ghirlanda} G.,  {Salvaterra} R.,  {Burlon} D.,  {Campana} S.,  {Melandri} A.,
  {Bernardini} M.~G.,  {Covino} S.,  {D'Avanzo} P.,  {D'Elia} V.,  {Ghisellini}
  G.,  {Nava} L.,  {Prandoni} I.,  {Sironi} L.,  {Tagliaferri} G.,  {Vergani}
  S.~D.,    {Wolter} A.,  2013, MNRAS, 435, 2543

\bibitem[\protect\citeauthoryear{{Granot} \& {Sari}}{{Granot} \&
  {Sari}}{2002}]{granot2002}
{Granot} J.,  {Sari} R.,  2002, ApJ, 568, 820

\bibitem[\protect\citeauthoryear{{Harry} \& {LIGO Scientific
  Collaboration}}{{Harry} \& {LIGO Scientific Collaboration}}{2010}]{harry2010}
{Harry} G.~M.,  {LIGO Scientific Collaboration} 2010, Classical and Quantum
  Gravity, 27, 084006

\bibitem[\protect\citeauthoryear{Helmstrom}{Helmstrom}{1968}]{helmstrom-1968}
Helmstrom C.~W.,  1968, Statistical Theory of Signal Detection, 2nd edition.
Pergamon Press, London

\bibitem[\protect\citeauthoryear{{Hogg}}{{Hogg}}{1999}]{hogg1999}
{Hogg} D.~W.,  1999, arXiv: astro-ph/9905116

\bibitem[\protect\citeauthoryear{{Jaeger}, {Hyman}, {Kassim} \&
  {Lazio}}{{Jaeger} et~al.}{2012}]{jaeger2012}
{Jaeger} T.~R.,  {Hyman} S.~D.,  {Kassim} N.~E.,    {Lazio} T.~J.~W.,  2012,
  AJ, 143, 96

\bibitem[\protect\citeauthoryear{{Kelley}, {Mandel} \& {Ramirez-Ruiz}}{{Kelley}
  et~al.}{2013}]{kelley2013}
{Kelley} L.~Z.,  {Mandel} I.,    {Ramirez-Ruiz} E.,  2013, Phys. Rev. D, 87,
  123004

\bibitem[\protect\citeauthoryear{{Kochanek} \& {Piran}}{{Kochanek} \&
  {Piran}}{1993}]{kochanek1993}
{Kochanek} C.~S.,  {Piran} T.,  1993, ApJ, 417, L17

\bibitem[\protect\citeauthoryear{{Lazio}, {Clarke}, {Lane}, {Gross}, {Kassim},
  {Ray}, {Wood}, {York}, {Kerkhoff}, {Hicks}, {Polisensky}, {Stewart},
  {Paravastu Dalal}, {Cohen} \& {Erickson}}{{Lazio} et~al.}{2010}]{lazio2010}
{Lazio} T.~J.~W.,  {Clarke} T.~E.,  {Lane} W.~M.,  {Gross} C.,  {Kassim} N.~E.,
   {Ray} P.~S.,  {Wood} D.,  {York} J.~A.,  {Kerkhoff} A.,  {Hicks} B.,
  {Polisensky} E.,  {Stewart} K.,  {Paravastu Dalal} N.,  {Cohen} A.~S.,
  {Erickson} W.~C.,  2010, AJ, 140, 1995

\bibitem[\protect\citeauthoryear{{Levinson}, {Ofek}, {Waxman} \&
  {Gal-Yam}}{{Levinson} et~al.}{2002}]{levinson2002}
{Levinson} A.,  {Ofek} E.~O.,  {Waxman} E.,    {Gal-Yam} A.,  2002, ApJ, 576,
  923

\bibitem[\protect\citeauthoryear{{Li} \& {Paczy{\'n}ski}}{{Li} \&
  {Paczy{\'n}ski}}{1998}]{li1998}
{Li} L.-X.,  {Paczy{\'n}ski} B.,  1998, ApJ, 507, L59

\bibitem[\protect\citeauthoryear{{LIGO Scientific Collaboration}, {Virgo
  Collaboration}, {Aasi}, {Abadie}, {Abbott}, {Abbott}, {Abbott}, {Abernathy},
  {Accadia}, {Acernese} \& et al.}{{LIGO Scientific Collaboration}
  et~al.}{2013}]{ligo2013}
{LIGO Scientific Collaboration} {Virgo Collaboration} {Aasi} J.,  {Abadie} J.,
  {Abbott} B.~P.,  {Abbott} R.,  {Abbott} T.~D.,  {Abernathy} M.,  {Accadia}
  T.,  {Acernese} F.,    et al. 2013, arXiv: 1304.0670

\bibitem[\protect\citeauthoryear{{LIGO Scientific Collaboration}, {Virgo
  Collaboration}, {Abadie}, {Abbott}, {Abbott}, {Abbott}, {Abernathy},
  {Accadia}, {Acernese}, {Adams} \& et al.}{{LIGO Scientific Collaboration}
  et~al.}{2012}]{ligo2012}
{LIGO Scientific Collaboration} {Virgo Collaboration} {Abadie} J.,  {Abbott}
  B.~P.,  {Abbott} R.,  {Abbott} T.~D.,  {Abernathy} M.,  {Accadia} T.,
  {Acernese} F.,  {Adams} C.,    et al. 2012, A\&A, 539, A124

\bibitem[\protect\citeauthoryear{{Metzger} \& {Berger}}{{Metzger} \&
  {Berger}}{2012}]{metzger2012}
{Metzger} B.~D.,  {Berger} E.,  2012, ApJ, 746, 48

\bibitem[\protect\citeauthoryear{{Metzger}, {Mart{\'{\i}}nez-Pinedo}, {Darbha},
  {Quataert}, {Arcones}, {Kasen}, {Thomas}, {Nugent}, {Panov} \&
  {Zinner}}{{Metzger} et~al.}{2010}]{metzger2010}
{Metzger} B.~D.,  {Mart{\'{\i}}nez-Pinedo} G.,  {Darbha} S.,  {Quataert} E.,
  {Arcones} A.,  {Kasen} D.,  {Thomas} R.,  {Nugent} P.,  {Panov} I.~V.,
  {Zinner} N.~T.,  2010, MNRAS, 406, 2650

\bibitem[\protect\citeauthoryear{{Moniez}}{{Moniez}}{2012}]{Moniez:2012iy}
{Moniez} M.,  2012, arXiv: 1208.6427

\bibitem[\protect\citeauthoryear{{Murphy}, {Chatterjee}, {Kaplan}, {Banyer},
  {Bell}, {Bignall}, {Bower}, {Cameron}, {Coward}, {Cordes} \& et al.}{{Murphy}
  et~al.}{2013}]{murphy2013}
{Murphy} T.,  {Chatterjee} S.,  {Kaplan} D.~L.,  {Banyer} J.,  {Bell} M.~E.,
  {Bignall} H.~E.,  {Bower} G.~C.,  {Cameron} R.~A.,  {Coward} D.~M.,  {Cordes}
  J.~M.,    et al. 2013, PASA, 30, 6

\bibitem[\protect\citeauthoryear{{Nakar}}{{Nakar}}{2007}]{nakar2007}
{Nakar} E.,  2007, Physics Reports, 442, 166

\bibitem[\protect\citeauthoryear{{Nakar} \& {Piran}}{{Nakar} \&
  {Piran}}{2011}]{nakar2011}
{Nakar} E.,  {Piran} T.,  2011, Nature, 478, 82

\bibitem[\protect\citeauthoryear{{Narayan}, {Paczynski} \& {Piran}}{{Narayan}
  et~al.}{1992}]{narayan1992}
{Narayan} R.,  {Paczynski} B.,    {Piran} T.,  1992, ApJ, 395, L83

\bibitem[\protect\citeauthoryear{Neyman \& Pearson}{Neyman \&
  Pearson}{1933}]{neyman-1933}
Neyman J.,  Pearson E.~S.,  1933, Philosophical Transactions of the Royal
  Society of London. Series A, Containing Papers of a Mathematical or Physical
  Character, 231, 289

\bibitem[\protect\citeauthoryear{{Nissanke}, {Holz}, {Hughes}, {Dalal} \&
  {Sievers}}{{Nissanke} et~al.}{2010}]{nissanke2010}
{Nissanke} S.,  {Holz} D.~E.,  {Hughes} S.~A.,  {Dalal} N.,    {Sievers} J.~L.,
   2010, ApJ, 725, 496

\bibitem[\protect\citeauthoryear{{Panaitescu}}{{Panaitescu}}{2006}]{panaitescu%
2006}
{Panaitescu} A.,  2006, MNRAS, 367, L42

\bibitem[\protect\citeauthoryear{{Perna} \& {Belczynski}}{{Perna} \&
  {Belczynski}}{2002}]{perna2002}
{Perna} R.,  {Belczynski} K.,  2002, ApJ, 570, 252

\bibitem[\protect\citeauthoryear{{Phinney}}{{Phinney}}{2009}]{phinney2009}
{Phinney} E.~S.,  2009, in astro2010: The Astronomy and Astrophysics Decadal
  Survey Vol.~2010 of Astronomy, {Finding and Using Electromagnetic
  Counterparts of Gravitational Wave Sources}.
p.~235

\bibitem[\protect\citeauthoryear{{Piran}, {Nakar} \& {Rosswog}}{{Piran}
  et~al.}{2013}]{piran2013}
{Piran} T.,  {Nakar} E.,    {Rosswog} S.,  2013, MNRAS, 430, 2121

\bibitem[\protect\citeauthoryear{{Planck Collaboration}, {Ade}, {Aghanim},
  {Armitage-Caplan}, {Arnaud}, {Ashdown}, {Atrio-Barandela}, {Aumont},
  {Baccigalupi}, {Banday} \& et al.}{{Planck Collaboration}
  et~al.}{2013}]{planck2013}
{Planck Collaboration} {Ade} P.~A.~R.,  {Aghanim} N.,  {Armitage-Caplan} C.,
  {Arnaud} M.,  {Ashdown} M.,  {Atrio-Barandela} F.,  {Aumont} J.,
  {Baccigalupi} C.,  {Banday} A.~J.,    et al. 2013, arXiv: 1303.5076

\bibitem[\protect\citeauthoryear{{Sari}, {Piran} \& {Narayan}}{{Sari}
  et~al.}{1998}]{sari1998}
{Sari} R.,  {Piran} T.,    {Narayan} R.,  1998, ApJ, 497, L17

\bibitem[\protect\citeauthoryear{{Schutz}}{{Schutz}}{1986}]{schutz1986}
{Schutz} B.~F.,  1986, Nature, 323, 310

\bibitem[\protect\citeauthoryear{{Siellez}, {Boer} \& {Gendre}}{{Siellez}
  et~al.}{2013}]{siellez2013}
{Siellez} K.,  {Boer} M.,    {Gendre} B.,  2013, arXiv: 1310.2106

\bibitem[\protect\citeauthoryear{{Singer}, {Cenko}, {Kasliwal}, {Perley},
  {Ofek}, {Brown}, {Nugent}, {Kulkarni}, {Corsi}, {Frail} \& et al.}{{Singer}
  et~al.}{2013}]{singer2013}
{Singer} L.~P.,  {Cenko} S.~B.,  {Kasliwal} M.~M.,  {Perley} D.~A.,  {Ofek}
  E.~O.,  {Brown} D.~A.,  {Nugent} P.~E.,  {Kulkarni} S.~R.,  {Corsi} A.,
  {Frail} D.~A.,    et al. 2013, ApJ, 776, L34

\bibitem[\protect\citeauthoryear{{Singer}, {Price}, {Farr}, {Urban}, {Pankow},
  {Vitale}, {Veitch}, {Farr}, {Hanna}, {Cannon} \& et al.}{{Singer}
  et~al.}{2014}]{singer2014}
{Singer} L.~P.,  {Price} L.~R.,  {Farr} B.,  {Urban} A.~L.,  {Pankow} C.,
  {Vitale} S.,  {Veitch} J.,  {Farr} W.~M.,  {Hanna} C.,  {Cannon} K.,    et
  al. 2014, arXiv: 1404.5623

\bibitem[\protect\citeauthoryear{{Sironi} \& {Giannios}}{{Sironi} \&
  {Giannios}}{2013}]{sironi2013}
{Sironi} L.,  {Giannios} D.,  2013, ApJ, 778, 107

\bibitem[\protect\citeauthoryear{{Soderberg}, {Berger}, {Kasliwal}, {Frail},
  {Price}, {Schmidt}, {Kulkarni}, {Fox}, {Cenko}, {Gal-Yam}, {Nakar} \&
  {Roth}}{{Soderberg} et~al.}{2006}]{soderberg2006}
{Soderberg} A.~M.,  {Berger} E.,  {Kasliwal} M.,  {Frail} D.~A.,  {Price}
  P.~A.,  {Schmidt} B.~P.,  {Kulkarni} S.~R.,  {Fox} D.~B.,  {Cenko} S.~B.,
  {Gal-Yam} A.,  {Nakar} E.,    {Roth} K.~C.,  2006, ApJ, 650, 261

\bibitem[\protect\citeauthoryear{{Tanvir}, {Levan}, {Fruchter}, {Hjorth},
  {Hounsell}, {Wiersema} \& {Tunnicliffe}}{{Tanvir} et~al.}{2013}]{tanvir2013}
{Tanvir} N.~R.,  {Levan} A.~J.,  {Fruchter} A.~S.,  {Hjorth} J.,  {Hounsell}
  R.~A.,  {Wiersema} K.,    {Tunnicliffe} R.~L.,  2013, Nature, 500, 547

\bibitem[\protect\citeauthoryear{{The~Virgo~Collaboration}}{{The~Virgo~Collabo%
ration}}{2009}]{virgo2009}
{The~Virgo~Collaboration}, 2009, {Advanced Virgo Baseline Design}

\bibitem[\protect\citeauthoryear{{Tingay}, {Goeke}, {Bowman}, {Emrich}, {Ord},
  {Mitchell}, {Morales}, {Booler}, {Crosse}, {Wayth} \& et al.}{{Tingay}
  et~al.}{2013}]{tingay2013}
{Tingay} S.~J.,  {Goeke} R.,  {Bowman} J.~D.,  {Emrich} D.,  {Ord} S.~M.,
  {Mitchell} D.~A.,  {Morales} M.~F.,  {Booler} T.,  {Crosse} B.,  {Wayth}
  R.~B.,    et al. 2013, PASA, 30, 7

\bibitem[\protect\citeauthoryear{{Trott}, {Wayth}, {Macquart} \&
  {Tingay}}{{Trott} et~al.}{2011}]{trott2011}
{Trott} C.~M.,  {Wayth} R.~B.,  {Macquart} J.-P.~R.,    {Tingay} S.~J.,  2011,
  ApJ, 731, 81

\bibitem[\protect\citeauthoryear{{van Eerten}, {van der Horst} \&
  {MacFadyen}}{{van Eerten} et~al.}{2012}]{vaneerten2012_boxfit}
{van Eerten} H.,  {van der Horst} A.,    {MacFadyen} A.,  2012, ApJ, 749, 44

\bibitem[\protect\citeauthoryear{{van Eerten} \& {MacFadyen}}{{van Eerten} \&
  {MacFadyen}}{2011}]{vaneerten2011}
{van Eerten} H.~J.,  {MacFadyen} A.~I.,  2011, ApJ, 733, L37

\bibitem[\protect\citeauthoryear{{van Eerten} \& {MacFadyen}}{{van Eerten} \&
  {MacFadyen}}{2012}]{vaneerten2012}
{van Eerten} H.~J.,  {MacFadyen} A.~I.,  2012, ApJ, 747, L30

\bibitem[\protect\citeauthoryear{{van Haarlem}, {Wise}, {Gunst}, {Heald},
  {McKean}, {Hessels}, {de Bruyn}, {Nijboer}, {Swinbank}, {Fallows} \& et
  al.}{{van Haarlem} et~al.}{2013}]{vanhaarlem2013}
{van Haarlem} M.~P.,  {Wise} M.~W.,  {Gunst} A.~W.,  {Heald} G.,  {McKean}
  J.~P.,  {Hessels} J.~W.~T.,  {de Bruyn} A.~G.,  {Nijboer} R.,  {Swinbank} J.,
   {Fallows} R.,    et al. 2013, A\&A, 556, A2

\bibitem[\protect\citeauthoryear{{Zhang}, {Kong}, {Huang}, {Li} \&
  {Li}}{{Zhang} et~al.}{2014}]{zhang2014}
{Zhang} Z.-B.,  {Kong} S.-W.,  {Huang} Y.-F.,  {Li} D.,    {Li} L.-B.,  2014,
  arXiv: 1402.6810

\end{thebibliography}

\label{lastpage}

\end{document}